\documentclass[10pt,conference]{IEEEtran}

\usepackage{amsmath,amssymb,amsfonts}
\usepackage{algorithmic}

\usepackage{fvextra}
\usepackage{textcomp}
\usepackage{xcolor}
\usepackage{minted}
\usepackage{subcaption}

\usepackage[normalem]{ulem}
\usepackage{balance}
\usepackage{listings}
\usepackage{xspace}
\usepackage{subcaption}
\usepackage{booktabs}
\usepackage{url}
\usepackage{framed}
\usepackage{forest}
\usepackage{tcolorbox} %
\usepackage{subcaption}
\usetikzlibrary{positioning, arrows.meta, shapes.arrows, shapes.misc, fit, backgrounds, calc}
\usepackage{algorithm}
\usepackage{enumitem}

\definecolor{Gray}{gray}{0.9}
\definecolor{cadmiumgreen}{rgb}{0.0, 0.42, 0.24}
\definecolor{Green}{rgb}{0.6,1,0.8}
\definecolor{Red}{rgb}{0.99, 0.76, 0.8}

\definecolor{dkgreen}{rgb}{0,0.6,0}
\definecolor{gray}{rgb}{0.5,0.5,0.5}
\definecolor{mauve}{rgb}{0.58,0,0.82}

\usepackage{numprint}
\npthousandsep{,}

\newcommand{\MyPara}[1]{\noindent\textbf{#1}:} %

\newcommand{\fscore}{F1 score}

\newcommand{\treevada}{\textsc{TreeVada}}

\newcommand{\kedavra}{\textsc{Kedavra}}
\newcommand{\crucio}{\textsc{Crucio}}
\newcommand{\arvada}{\textsc{Arvada}}
\newcommand{\glade}{\textsc{Glade}}
\newcommand{\gladeRep}{\textsc{Glade-II}}
\newcommand{\toolNamePlain}{XVada} %

\newcommand{\toolName}{\textsc{\toolNamePlain}}

\newcommand{\Tool}{\toolName}

\newcommand{\Accessed}{Accessed June 2026: }

\newcommand{\RQAccuracy}{RQ1}
\newcommand{\RQAccuracyFull}{%
How accurate are the grammars inferred by \Tool{} compared with those inferred by state-of-the-art approaches, when evaluated against the golden grammar of languages with varying sizes and complexities?
}

\newcommand{\RQCompact}{RQ2}
\newcommand{\RQCompactFull}{
How compact are \Tool{}-inferred grammars compared with similarly accurate grammars produced by state-of-the-art approaches?
}

\newcommand{\RQAblation}{RQ3}
\newcommand{\RQAblationFull}{How does HDD affect the accuracy of grammars inferred by \Tool{}?}

\newcommand{\RQFuzzing}{RQ4}
\newcommand{\RQFuzzingFull}{How effective are \Tool{}-inferred grammars for practical fuzzing tasks?}

\newcommand{\term}[1]{\textcolor{teal}{\texttt{#1}}}

\newcommand{\Def}[2]{\expandafter\newcommand\csname rmk-#1\endcsname{#2}}
\newcommand{\Use}[1]{\csname rmk-#1\endcsname}

\Def{turtle_grammar_stat_term}{10}
\Def{while_grammar_stat_term}{20}
\Def{lisp_grammar_stat_term}{7}
\Def{xml_grammar_stat_term}{17}
\Def{json_grammar_stat_term}{12}
\Def{tinyc_grammar_stat_term}{15}
\Def{c-500_grammar_stat_term}{15}
\Def{tiny_grammar_stat_term}{10}
\Def{minic_grammar_stat_term}{22}
\Def{curl_grammar_stat_term}{n/a}
\Def{liquid_grammar_stat_term}{n/a}
\Def{lua_grammar_stat_term}{61}
\Def{c_grammar_stat_term}{108}
\Def{mysql_grammar_stat_term}{1,028}
\Def{java_grammar_stat_term}{128}
\Def{cpp_grammar_stat_term}{129}
\Def{rust_grammar_stat_term}{117}
\Def{Avg-Small-Program_grammar_stat_term}{0}
\Def{Avg-Big-Program_grammar_stat_term}{0}
\Def{turtle_grammar_stat_non_term}{4}
\Def{while_grammar_stat_non_term}{4}
\Def{lisp_grammar_stat_non_term}{5}
\Def{xml_grammar_stat_non_term}{7}
\Def{json_grammar_stat_non_term}{7}
\Def{tinyc_grammar_stat_non_term}{9}
\Def{c-500_grammar_stat_non_term}{9}
\Def{tiny_grammar_stat_non_term}{13}
\Def{minic_grammar_stat_non_term}{14}
\Def{curl_grammar_stat_non_term}{n/a}
\Def{liquid_grammar_stat_non_term}{n/a}
\Def{lua_grammar_stat_non_term}{37}
\Def{c_grammar_stat_non_term}{87}
\Def{mysql_grammar_stat_non_term}{319}
\Def{java_grammar_stat_non_term}{126}
\Def{cpp_grammar_stat_non_term}{174}
\Def{rust_grammar_stat_non_term}{196}
\Def{Avg-Small-Program_grammar_stat_non_term}{0}
\Def{Avg-Big-Program_grammar_stat_non_term}{0}
\Def{turtle_grammar_rhs_length}{6.25}
\Def{while_grammar_rhs_length}{13.00}
\Def{lisp_grammar_rhs_length}{5.00}
\Def{xml_grammar_rhs_length}{6.86}
\Def{json_grammar_rhs_length}{4.29}
\Def{tinyc_grammar_rhs_length}{4.67}
\Def{c-500_grammar_rhs_length}{4.67}
\Def{tiny_grammar_rhs_length}{2.69}
\Def{minic_grammar_rhs_length}{6.36}
\Def{curl_grammar_rhs_length}{n/a}
\Def{liquid_grammar_rhs_length}{n/a}
\Def{lua_grammar_rhs_length}{5.87}
\Def{c_grammar_rhs_length}{6.03}
\Def{mysql_grammar_rhs_length}{n/a}
\Def{java_grammar_rhs_length}{6.31}
\Def{cpp_grammar_rhs_length}{5.48}
\Def{rust_grammar_rhs_length}{5.78}
\Def{Avg-Small-Program_grammar_rhs_length}{.00}
\Def{Avg-Big-Program_grammar_rhs_length}{.00}
\Def{turtle_grammar_NT_times_RHS}{25}
\Def{while_grammar_NT_times_RHS}{52}
\Def{lisp_grammar_NT_times_RHS}{25}
\Def{xml_grammar_NT_times_RHS}{48}
\Def{json_grammar_NT_times_RHS}{30}
\Def{tinyc_grammar_NT_times_RHS}{42}
\Def{c-500_grammar_NT_times_RHS}{42}
\Def{tiny_grammar_NT_times_RHS}{35}
\Def{minic_grammar_NT_times_RHS}{89}
\Def{curl_grammar_NT_times_RHS}{n/a}
\Def{lua_grammar_NT_times_RHS}{217}
\Def{c_grammar_NT_times_RHS}{525}
\Def{mysql_grammar_NT_times_RHS}{n/a}
\Def{java_grammar_NT_times_RHS}{795}
\Def{cpp_grammar_NT_times_RHS}{954}
\Def{rust_grammar_NT_times_RHS}{1133}
\Def{Avg-Small-Program_grammar_NT_times_RHS}{n/a}
\Def{Avg-Big-Program_grammar_NT_times_RHS}{n/a}
\Def{turtle_grammar_mcc}{1.50}
\Def{while_grammar_mcc}{2.50}
\Def{lisp_grammar_mcc}{1.40}
\Def{xml_grammar_mcc}{2.43}
\Def{json_grammar_mcc}{1.43}
\Def{tinyc_grammar_mcc}{1.44}
\Def{c-500_grammar_mcc}{1.44}
\Def{tiny_grammar_mcc}{.69}
\Def{minic_grammar_mcc}{2.00}
\Def{curl_grammar_mcc}{n/a}
\Def{liquid_grammar_mcc}{n/a}
\Def{lua_grammar_mcc}{2.35}
\Def{c_grammar_mcc}{2.747}
\Def{mysql_grammar_mcc}{n/a}
\Def{java_grammar_mcc}{3.095}
\Def{cpp_grammar_mcc}{2.91}
\Def{rust_grammar_mcc}{3.107}
\Def{Avg-Small-Program_grammar_mcc}{.00}
\Def{Avg-Big-Program_grammar_mcc}{.00}
\Def{turtle_grammar_ntd}{4}
\Def{while_grammar_ntd}{3}
\Def{lisp_grammar_ntd}{4}
\Def{xml_grammar_ntd}{4}
\Def{json_grammar_ntd}{5}
\Def{tinyc_grammar_ntd}{8}
\Def{tiny_grammar_ntd}{7}
\Def{minic_grammar_ntd}{7}
\Def{curl_grammar_ntd}{n/a}
\Def{liquid_grammar_ntd}{n/a}
\Def{lua_grammar_ntd}{7}
\Def{c_grammar_ntd}{13}
\Def{mysql_grammar_ntd}{15}
\Def{java_grammar_ntd}{10}
\Def{cpp_grammar_ntd}{14}
\Def{rust_grammar_ntd}{13}
\Def{Avg-Small-Program_grammar_ntd}{n/a}
\Def{Avg-Big-Program_grammar_ntd}{n/a}
\Def{turtle_grammar_lrs}{66}
\Def{while_grammar_lrs}{124}
\Def{lisp_grammar_lrs}{84}
\Def{xml_grammar_lrs}{342}
\Def{json_grammar_lrs}{85}
\Def{tinyc_grammar_lrs}{265}
\Def{c-500_grammar_lrs}{265}
\Def{tiny_grammar_lrs}{64}
\Def{minic_grammar_lrs}{355}
\Def{curl_grammar_lrs}{n/a}
\Def{liquid_grammar_lrs}{n/a}
\Def{lua_grammar_lrs}{2,868}
\Def{c_grammar_lrs}{4,675}
\Def{mysql_grammar_lrs}{n/a}
\Def{java_grammar_lrs}{8,316}
\Def{cpp_grammar_lrs}{27,213}
\Def{rust_grammar_lrs}{21,427}
\Def{Avg-Small-Program_grammar_lrs}{0}
\Def{Avg-Big-Program_grammar_lrs}{0}
\Def{turtle_grammar_lat}{8.64}
\Def{while_grammar_lat}{10.14}
\Def{lisp_grammar_lat}{17.12}
\Def{xml_grammar_lat}{31.00}
\Def{json_grammar_lat}{13.23}
\Def{tinyc_grammar_lat}{59.69}
\Def{c-500_grammar_lat}{59.69}
\Def{tiny_grammar_lat}{27.10}
\Def{minic_grammar_lat}{68.04}
\Def{curl_grammar_lat}{n/a}
\Def{liquid_grammar_lat}{n/a}
\Def{lua_grammar_lat}{991.84}
\Def{c_grammar_lat}{647.93}
\Def{mysql_grammar_lat}{n/a}
\Def{java_grammar_lat}{1649.96}
\Def{cpp_grammar_lat}{5150.415}
\Def{rust_grammar_lat}{3153.229}
\Def{Avg-Small-Program_grammar_lat}{.00}
\Def{Avg-Big-Program_grammar_lat}{.00}
\Def{turtle_grammar_ltpsm}{8}
\Def{while_grammar_ltpsm}{16}
\Def{lisp_grammar_ltpsm}{5}
\Def{xml_grammar_ltpsm}{8}
\Def{json_grammar_ltpsm}{8}
\Def{tinyc_grammar_ltpsm}{12}
\Def{c-500_grammar_ltpsm}{12}
\Def{tiny_grammar_ltpsm}{8}
\Def{minic_grammar_ltpsm}{17}
\Def{curl_grammar_ltpsm}{n/a}
\Def{liquid_grammar_ltpsm}{n/a}
\Def{lua_grammar_ltpsm}{52}
\Def{c_grammar_ltpsm}{106}
\Def{mysql_grammar_ltpsm}{n/a}
\Def{java_grammar_ltpsm}{121}
\Def{cpp_grammar_ltpsm}{126}
\Def{rust_grammar_ltpsm}{117}
\Def{Avg-Small-Program_grammar_ltpsm}{0}
\Def{Avg-Big-Program_grammar_ltpsm}{0}
\Def{turtle_program_seed_no}{35}
\Def{while_program_seed_no}{30}
\Def{lisp_program_seed_no}{30}
\Def{xml_program_seed_no}{20}
\Def{json_program_seed_no}{30}
\Def{tinyc_program_seed_no}{25}
\Def{c-500_program_seed_no}{10}
\Def{tiny_program_seed_no}{10}
\Def{minic_program_seed_no}{10}
\Def{curl_program_seed_no}{25}
\Def{liquid_program_seed_no}{50}
\Def{c_program_seed_no}{60}
\Def{lua_program_seed_no}{50}
\Def{mysql_program_seed_no}{60}
\Def{java_program_seed_no}{50}
\Def{cpp_program_seed_no}{60}
\Def{rust_program_seed_no}{50}
\Def{Avg-Small-Program_program_seed_no}{0}
\Def{Avg-Big-Program_program_seed_no}{0}
\Def{turtle_avg_char}{41.09}
\Def{while_avg_char}{171.43}
\Def{lisp_avg_char}{79.20}
\Def{xml_avg_char}{27.80}
\Def{json_avg_char}{11.73}
\Def{tinyc_avg_char}{96.16}
\Def{c-500_avg_char}{514.00}
\Def{tiny_avg_char}{39.70}
\Def{minic_avg_char}{107.00}
\Def{curl_avg_char}{22.08}
\Def{liquid_avg_char}{54.5}
\Def{c_avg_char}{170.93}
\Def{lua_avg_char}{95.34}
\Def{mysql_avg_char}{104.23}
\Def{java_avg_char}{120.18}
\Def{cpp_avg_char}{248.58}
\Def{rust_avg_char}{188.76}
\Def{Avg-Small-Program_avg_char}{.00}
\Def{Avg-Big-Program_avg_char}{.00}
\Def{turtle_max_char}{126}
\Def{while_max_char}{793}
\Def{lisp_max_char}{428}
\Def{xml_max_char}{78}
\Def{json_max_char}{90}
\Def{tinyc_max_char}{257}
\Def{c-500_max_char}{593}
\Def{tiny_max_char}{76}
\Def{minic_max_char}{368}
\Def{curl_max_char}{39}
\Def{liquid_max_char}{171}
\Def{c_max_char}{523}
\Def{lua_max_char}{838}
\Def{mysql_max_char}{517}
\Def{java_max_char}{660}
\Def{cpp_max_char}{1,146}
\Def{rust_max_char}{1,074}
\Def{Avg-Small-Program_max_char}{0}
\Def{Avg-Big-Program_max_char}{0}
\Def{turtle_avg_tokens}{27.00}
\Def{while_avg_tokens}{110.70}
\Def{lisp_avg_tokens}{75.50}
\Def{xml_avg_tokens}{23.20}
\Def{json_avg_tokens}{8.20}
\Def{tinyc_avg_tokens}{47.24}
\Def{c-500_avg_tokens}{244.30}
\Def{tiny_avg_tokens}{15.50}
\Def{minic_avg_tokens}{45.70}
\Def{curl_avg_tokens}{14.80}
\Def{liquid_avg_tokens}{22.94}
\Def{c_avg_tokens}{63.6}
\Def{lua_avg_tokens}{38.54}
\Def{mysql_avg_tokens}{38.9}
\Def{java_avg_tokens}{34.52}
\Def{cpp_avg_tokens}{91.18}
\Def{rust_avg_tokens}{74.42}
\Def{Avg-Small-Program_avg_tokens}{.00}
\Def{Avg-Big-Program_avg_tokens}{.00}
\Def{turtle_max_tokens}{82}
\Def{while_max_tokens}{534}
\Def{lisp_max_tokens}{411}
\Def{xml_max_tokens}{60}
\Def{json_max_tokens}{72}
\Def{tinyc_max_tokens}{126}
\Def{c-500_max_tokens}{279}
\Def{tiny_max_tokens}{31}
\Def{minic_max_tokens}{158}
\Def{curl_max_tokens}{25}
\Def{liquid_max_tokens}{84}
\Def{c_max_tokens}{227}
\Def{lua_max_tokens}{334}
\Def{mysql_max_tokens}{215}
\Def{java_max_tokens}{217}
\Def{cpp_max_tokens}{427}
\Def{rust_max_tokens}{444}
\Def{Avg-Small-Program_max_tokens}{0}
\Def{Avg-Big-Program_max_tokens}{0}
\Def{turtle_branching_factor}{8.28}
\Def{while_branching_factor}{24.39}
\Def{lisp_branching_factor}{4.08}
\Def{xml_branching_factor}{23.20}
\Def{json_branching_factor}{3.77}
\Def{tinyc_branching_factor}{7.71}
\Def{c-500_branching_factor}{8.45}
\Def{tiny_branching_factor}{15.50}
\Def{minic_branching_factor}{5.33}
\Def{curl_branching_factor}{14.80}
\Def{liquid_branching_factor}{6.82}
\Def{c_branching_factor}{8.14}
\Def{lua_branching_factor}{19.284}
\Def{mysql_branching_factor}{25.44}
\Def{java_branching_factor}{10.779}
\Def{cpp_branching_factor}{6.44}
\Def{rust_branching_factor}{38.85}
\Def{Avg-Small-Program_branching_factor}{.00}
\Def{Avg-Big-Program_branching_factor}{.00}
\Def{turtle_ID}{M1}
\Def{turtle_programname}{turtle}
\Def{turtle_Kedavra_precision}{1.00}
\Def{turtle_Kedavra_recall}{1.00}
\Def{turtle_Kedavra_f1}{1.00}
\Def{turtle_Kedavra_build_time}{96}
\Def{turtle_treevada_precision}{.95}
\Def{turtle_treevada_recall}{1.00}
\Def{turtle_treevada_f1}{.97}
\Def{turtle_treevada_build_time}{94}
\Def{turtle_crucio_precision}{1.00}
\Def{turtle_crucio_recall}{.01}
\Def{turtle_crucio_f1}{.01}
\Def{turtle_crucio_build_time}{20}
\Def{turtle_xvada_precision}{1.00}
\Def{turtle_xvada_recall}{1.00}
\Def{turtle_xvada_f1}{1.00}
\Def{turtle_xvada_build_time}{92}
\Def{while_ID}{M2}
\Def{while_programname}{while}
\Def{while_Kedavra_precision}{1.00}
\Def{while_Kedavra_recall}{1.00}
\Def{while_Kedavra_f1}{1.00}
\Def{while_Kedavra_build_time}{513}
\Def{while_treevada_precision}{1.00}
\Def{while_treevada_recall}{1.00}
\Def{while_treevada_f1}{1.00}
\Def{while_treevada_build_time}{1,682}
\Def{while_crucio_precision}{1.00}
\Def{while_crucio_recall}{.31}
\Def{while_crucio_f1}{.47}
\Def{while_crucio_build_time}{1,726}
\Def{while_xvada_precision}{1.00}
\Def{while_xvada_recall}{1.00}
\Def{while_xvada_f1}{1.00}
\Def{while_xvada_build_time}{840}
\Def{lisp_ID}{M3}
\Def{lisp_programname}{lisp}
\Def{lisp_Kedavra_precision}{1.00}
\Def{lisp_Kedavra_recall}{1.00}
\Def{lisp_Kedavra_f1}{1.00}
\Def{lisp_Kedavra_build_time}{192}
\Def{lisp_treevada_precision}{.98}
\Def{lisp_treevada_recall}{1.00}
\Def{lisp_treevada_f1}{.99}
\Def{lisp_treevada_build_time}{813}
\Def{lisp_crucio_precision}{1.00}
\Def{lisp_crucio_recall}{.00}
\Def{lisp_crucio_f1}{.01}
\Def{lisp_crucio_build_time}{28}
\Def{lisp_xvada_precision}{1.00}
\Def{lisp_xvada_recall}{1.00}
\Def{lisp_xvada_f1}{1.00}
\Def{lisp_xvada_build_time}{435}
\Def{xml_ID}{M4}
\Def{xml_programname}{xml}
\Def{xml_Kedavra_precision}{1.00}
\Def{xml_Kedavra_recall}{1.00}
\Def{xml_Kedavra_f1}{1.00}
\Def{xml_Kedavra_build_time}{366}
\Def{xml_treevada_precision}{1.00}
\Def{xml_treevada_recall}{1.00}
\Def{xml_treevada_f1}{1.00}
\Def{xml_treevada_build_time}{128}
\Def{xml_crucio_precision}{.23}
\Def{xml_crucio_recall}{.06}
\Def{xml_crucio_f1}{.09}
\Def{xml_crucio_build_time}{229}
\Def{xml_xvada_precision}{1.00}
\Def{xml_xvada_recall}{.87}
\Def{xml_xvada_f1}{.93}
\Def{xml_xvada_build_time}{375}
\Def{json_ID}{M5}
\Def{json_programname}{json}
\Def{json_Kedavra_precision}{1.00}
\Def{json_Kedavra_recall}{.97}
\Def{json_Kedavra_f1}{.98}
\Def{json_Kedavra_build_time}{38}
\Def{json_treevada_precision}{.96}
\Def{json_treevada_recall}{.90}
\Def{json_treevada_f1}{.93}
\Def{json_treevada_build_time}{26}
\Def{json_crucio_precision}{.14}
\Def{json_crucio_recall}{.02}
\Def{json_crucio_f1}{.03}
\Def{json_crucio_build_time}{47}
\Def{json_xvada_precision}{1.00}
\Def{json_xvada_recall}{.89}
\Def{json_xvada_f1}{.94}
\Def{json_xvada_build_time}{59}
\Def{tinyc_ID}{M6}
\Def{tinyc_programname}{tinyc}
\Def{tinyc_Kedavra_precision}{1.00}
\Def{tinyc_Kedavra_recall}{1.00}
\Def{tinyc_Kedavra_f1}{1.00}
\Def{tinyc_Kedavra_build_time}{116}
\Def{tinyc_treevada_precision}{.86}
\Def{tinyc_treevada_recall}{.97}
\Def{tinyc_treevada_f1}{.91}
\Def{tinyc_treevada_build_time}{2,237}
\Def{tinyc_crucio_precision}{.48}
\Def{tinyc_crucio_recall}{.00}
\Def{tinyc_crucio_f1}{.00}
\Def{tinyc_crucio_build_time}{36}
\Def{tinyc_xvada_precision}{1.00}
\Def{tinyc_xvada_recall}{.80}
\Def{tinyc_xvada_f1}{.89}
\Def{tinyc_xvada_build_time}{521}
\Def{c-500_ID}{M7}
\Def{c-500_programname}{c-500}
\Def{c-500_Kedavra_precision}{1.00}
\Def{c-500_Kedavra_recall}{1.00}
\Def{c-500_Kedavra_f1}{1.00}
\Def{c-500_Kedavra_build_time}{224}
\Def{c-500_treevada_precision}{.93}
\Def{c-500_treevada_recall}{.53}
\Def{c-500_treevada_f1}{.67}
\Def{c-500_treevada_build_time}{5,085}
\Def{c-500_crucio_precision}{.48}
\Def{c-500_crucio_recall}{.00}
\Def{c-500_crucio_f1}{.00}
\Def{c-500_crucio_build_time}{35}
\Def{c-500_xvada_precision}{.83}
\Def{c-500_xvada_recall}{.83}
\Def{c-500_xvada_f1}{.83}
\Def{c-500_xvada_build_time}{1,489}
\Def{tiny_ID}{M8}
\Def{tiny_programname}{tiny}
\Def{tiny_Kedavra_precision}{1.00}
\Def{tiny_Kedavra_recall}{.50}
\Def{tiny_Kedavra_f1}{.67}
\Def{tiny_Kedavra_build_time}{26}
\Def{tiny_treevada_precision}{.95}
\Def{tiny_treevada_recall}{.68}
\Def{tiny_treevada_f1}{.79}
\Def{tiny_treevada_build_time}{104}
\Def{tiny_crucio_precision}{.53}
\Def{tiny_crucio_recall}{.00}
\Def{tiny_crucio_f1}{.00}
\Def{tiny_crucio_build_time}{62}
\Def{tiny_xvada_precision}{.89}
\Def{tiny_xvada_recall}{.72}
\Def{tiny_xvada_f1}{.80}
\Def{tiny_xvada_build_time}{81}
\Def{minic_ID}{M9}
\Def{minic_programname}{minic}
\Def{minic_Kedavra_precision}{1.00}
\Def{minic_Kedavra_recall}{.48}
\Def{minic_Kedavra_f1}{.65}
\Def{minic_Kedavra_build_time}{246}
\Def{minic_treevada_precision}{.60}
\Def{minic_treevada_recall}{.62}
\Def{minic_treevada_f1}{.61}
\Def{minic_treevada_build_time}{1,033}
\Def{minic_crucio_precision}{.95}
\Def{minic_crucio_recall}{.00}
\Def{minic_crucio_f1}{.00}
\Def{minic_crucio_build_time}{194}
\Def{minic_xvada_precision}{.96}
\Def{minic_xvada_recall}{.59}
\Def{minic_xvada_f1}{.73}
\Def{minic_xvada_build_time}{588}
\Def{curl_ID}{M10}
\Def{curl_programname}{curl}
\Def{curl_Kedavra_precision}{.96}
\Def{curl_Kedavra_recall}{.14}
\Def{curl_Kedavra_f1}{.25}
\Def{curl_Kedavra_build_time}{159}
\Def{curl_treevada_precision}{.94}
\Def{curl_treevada_recall}{.68}
\Def{curl_treevada_f1}{.79}
\Def{curl_treevada_build_time}{135}
\Def{curl_crucio_precision}{.72}
\Def{curl_crucio_recall}{.03}
\Def{curl_crucio_f1}{.05}
\Def{curl_crucio_build_time}{19,099}
\Def{curl_xvada_precision}{.57}
\Def{curl_xvada_recall}{1.00}
\Def{curl_xvada_f1}{.73}
\Def{curl_xvada_build_time}{366}
\Def{Avg_lo_ID}{M11}
\Def{Avg_lo_programname}{Lo\textsubscript{avg}}
\Def{Avg_lo_Kedavra_precision}{1.00}
\Def{Avg_lo_Kedavra_recall}{.81}
\Def{Avg_lo_Kedavra_f1}{.85}
\Def{Avg_lo_Kedavra_build_time}{198}
\Def{Avg_lo_treevada_precision}{.92}
\Def{Avg_lo_treevada_recall}{.84}
\Def{Avg_lo_treevada_f1}{.87}
\Def{Avg_lo_treevada_build_time}{1,134}
\Def{Avg_lo_crucio_precision}{.65}
\Def{Avg_lo_crucio_recall}{.04}
\Def{Avg_lo_crucio_f1}{.07}
\Def{Avg_lo_crucio_build_time}{2,148}
\Def{Avg_lo_xvada_precision}{.92}
\Def{Avg_lo_xvada_recall}{.87}
\Def{Avg_lo_xvada_f1}{.88}
\Def{Avg_lo_xvada_build_time}{485}
\Def{lua_ID}{M12}
\Def{liquid_programname}{liquid}
\Def{liquid_Kedavra_precision}{}
\Def{liquid_Kedavra_recall}{}
\Def{liquid_Kedavra_f1}{}
\Def{liquid_Kedavra_build_time}{crash}
\Def{liquid_treevada_precision}{.56}
\Def{liquid_treevada_recall}{.17}
\Def{liquid_treevada_f1}{.26}
\Def{liquid_treevada_build_time}{305}
\Def{liquid_crucio_precision}{.51}
\Def{liquid_crucio_recall}{.00}
\Def{liquid_crucio_f1}{.00}
\Def{liquid_crucio_build_time}{10}
\Def{liquid_xvada_precision}{.98}
\Def{liquid_xvada_recall}{.98}
\Def{liquid_xvada_f1}{.99}
\Def{liquid_xvada_build_time}{847}
\Def{lua_programname}{lua}
\Def{lua_Kedavra_precision}{}
\Def{lua_Kedavra_recall}{}
\Def{lua_Kedavra_f1}{}
\Def{lua_Kedavra_build_time}{t/o}
\Def{lua_treevada_precision}{.81}
\Def{lua_treevada_recall}{.03}
\Def{lua_treevada_f1}{.06}
\Def{lua_treevada_build_time}{7,685}
\Def{lua_crucio_precision}{.39}
\Def{lua_crucio_recall}{.00}
\Def{lua_crucio_f1}{.00}
\Def{lua_crucio_build_time}{2,606}
\Def{lua_xvada_precision}{.50}
\Def{lua_xvada_recall}{.30}
\Def{lua_xvada_f1}{.37}
\Def{lua_xvada_build_time}{10,088}
\Def{c_ID}{M13}
\Def{c_programname}{c}
\Def{c_Kedavra_precision}{}
\Def{c_Kedavra_recall}{}
\Def{c_Kedavra_f1}{}
\Def{c_Kedavra_build_time}{crash}
\Def{c_treevada_precision}{.58}
\Def{c_treevada_recall}{.02}
\Def{c_treevada_f1}{.04}
\Def{c_treevada_build_time}{65,417}
\Def{c_crucio_precision}{.17}
\Def{c_crucio_recall}{.00}
\Def{c_crucio_f1}{.00}
\Def{c_crucio_build_time}{37,938}
\Def{c_xvada_precision}{.23}
\Def{c_xvada_recall}{.26}
\Def{c_xvada_f1}{.24}
\Def{c_xvada_build_time}{51,260}
\Def{java_ID}{M14}
\Def{java_programname}{java}
\Def{java_Kedavra_precision}{}
\Def{java_Kedavra_recall}{}
\Def{java_Kedavra_f1}{}
\Def{java_Kedavra_build_time}{t/o}
\Def{java_treevada_precision}{.96}
\Def{java_treevada_recall}{.05}
\Def{java_treevada_f1}{.10}
\Def{java_treevada_build_time}{11,254}
\Def{java_crucio_precision}{.00}
\Def{java_crucio_recall}{.00}
\Def{java_crucio_f1}{.00}
\Def{java_crucio_build_time}{15,466}
\Def{java_xvada_precision}{.84}
\Def{java_xvada_recall}{.62}
\Def{java_xvada_f1}{.71}
\Def{java_xvada_build_time}{11,060}
\Def{cpp_ID}{M15}
\Def{cpp_programname}{cpp}
\Def{cpp_Kedavra_precision}{}
\Def{cpp_Kedavra_recall}{}
\Def{cpp_Kedavra_f1}{}
\Def{cpp_Kedavra_build_time}{t/o}
\Def{cpp_treevada_precision}{.71}
\Def{cpp_treevada_recall}{.03}
\Def{cpp_treevada_f1}{.06}
\Def{cpp_treevada_build_time}{121,904}
\Def{cpp_crucio_precision}{}
\Def{cpp_crucio_recall}{}
\Def{cpp_crucio_f1}{}
\Def{cpp_crucio_build_time}{t/o}
\Def{cpp_xvada_precision}{.50}
\Def{cpp_xvada_recall}{.11}
\Def{cpp_xvada_f1}{.18}
\Def{cpp_xvada_build_time}{144,955}
\Def{rust_ID}{M16}
\Def{rust_programname}{rust}
\Def{rust_Kedavra_precision}{}
\Def{rust_Kedavra_recall}{}
\Def{rust_Kedavra_f1}{}
\Def{rust_Kedavra_build_time}{crash}
\Def{rust_treevada_precision}{.73}
\Def{rust_treevada_recall}{.10}
\Def{rust_treevada_f1}{.18}
\Def{rust_treevada_build_time}{72,122}
\Def{rust_crucio_precision}{}
\Def{rust_crucio_recall}{}
\Def{rust_crucio_f1}{}
\Def{rust_crucio_build_time}{t/o}
\Def{rust_xvada_precision}{.56}
\Def{rust_xvada_recall}{.60}
\Def{rust_xvada_f1}{.58}
\Def{rust_xvada_build_time}{62,477}
\Def{mysql_ID}{M17}
\Def{mysql_programname}{mysql}
\Def{mysql_Kedavra_precision}{}
\Def{mysql_Kedavra_recall}{}
\Def{mysql_Kedavra_f1}{}
\Def{mysql_Kedavra_build_time}{t/o}
\Def{mysql_treevada_precision}{.37}
\Def{mysql_treevada_recall}{.11}
\Def{mysql_treevada_f1}{.17}
\Def{mysql_treevada_build_time}{48,356}
\Def{mysql_crucio_precision}{}
\Def{mysql_crucio_recall}{}
\Def{mysql_crucio_f1}{}
\Def{mysql_crucio_build_time}{t/o}
\Def{mysql_xvada_precision}{.78}
\Def{mysql_xvada_recall}{.12}
\Def{mysql_xvada_f1}{.21}
\Def{mysql_xvada_build_time}{77,229}
\Def{Avg_hi_ID}{M18}
\Def{Avg_hi_programname}{Hi\textsubscript{avg}}
\Def{Avg_hi_Kedavra_precision}{n/a}
\Def{Avg_hi_Kedavra_recall}{n/a}
\Def{Avg_hi_Kedavra_f1}{n/a}
\Def{Avg_hi_Kedavra_build_time}{n/a}
\Def{Avg_hi_treevada_precision}{.68}
\Def{Avg_hi_treevada_recall}{.07}
\Def{Avg_hi_treevada_f1}{.12}
\Def{Avg_hi_treevada_build_time}{46,720}
\Def{Avg_hi_crucio_precision}{n/a}
\Def{Avg_hi_crucio_recall}{n/a}
\Def{Avg_hi_crucio_f1}{n/a}
\Def{Avg_hi_crucio_build_time}{n/a}
\Def{Avg_hi_xvada_precision}{.63}
\Def{Avg_hi_xvada_recall}{.43}
\Def{Avg_hi_xvada_f1}{.47}
\Def{Avg_hi_xvada_build_time}{51,131}
\Def{c_p_sample_and_std}{.02 $\pm$ 0.01}
\Def{c_p_swap_and_std}{.65 $\pm$ 0.04}
\Def{c_f1_sample_and_std}{.04 $\pm$ 0.02}
\Def{c_f1_swap_and_std}{.73 $\pm$ 0.04}
\Def{cpp_p_sample_and_std}{.07 $\pm$ 0.03}
\Def{cpp_p_swap_and_std}{.62 $\pm$ 0.02}
\Def{cpp_f1_sample_and_std}{.13 $\pm$ 0.05}
\Def{cpp_f1_swap_and_std}{.70 $\pm$ 0.01}
\Def{java_p_sample_and_std}{.05 $\pm$ 0.03}
\Def{java_p_swap_and_std}{.75 $\pm$ 0.04}
\Def{java_f1_sample_and_std}{.09 $\pm$ 0.05}
\Def{java_f1_swap_and_std}{.83 $\pm$ 0.03}
\Def{rust_p_sample_and_std}{.22 $\pm$ 0.05}
\Def{rust_p_swap_and_std}{.42 $\pm$ 0.04}
\Def{rust_f1_sample_and_std}{.34 $\pm$ 0.06}
\Def{rust_f1_swap_and_std}{.54 $\pm$ 0.03}
\Def{lua_p_sample_and_std}{.43 $\pm$ 0.05}
\Def{lua_p_swap_and_std}{.59 $\pm$ 0.05}
\Def{lua_f1_sample_and_std}{.56 $\pm$ 0.03}
\Def{lua_f1_swap_and_std}{.69 $\pm$ 0.04}
\Def{turtle_treevada_nt}{15}
\Def{turtle_treevada_T}{67}
\Def{turtle_treevada_A}{95}
\Def{turtle_treevada_IA}{1.31}
\Def{turtle_treevada_S}{124}
\Def{turtle_treevada_D}{5}
\Def{turtle_Kedavra_nt}{11}
\Def{turtle_Kedavra_T}{67}
\Def{turtle_Kedavra_A}{86}
\Def{turtle_Kedavra_IA}{1.21}
\Def{turtle_Kedavra_S}{104}
\Def{turtle_Kedavra_D}{5}
\Def{turtle_xvada_nt}{9}
\Def{turtle_xvada_T}{67}
\Def{turtle_xvada_A}{79}
\Def{turtle_xvada_IA}{1.27}
\Def{turtle_xvada_S}{100}
\Def{turtle_xvada_D}{4}
\Def{while_treevada_nt}{24}
\Def{while_treevada_T}{18}
\Def{while_treevada_A}{55}
\Def{while_treevada_IA}{2.05}
\Def{while_treevada_S}{113}
\Def{while_treevada_D}{6}
\Def{while_Kedavra_nt}{6}
\Def{while_Kedavra_T}{18}
\Def{while_Kedavra_A}{16}
\Def{while_Kedavra_IA}{3.35}
\Def{while_Kedavra_S}{54}
\Def{while_Kedavra_D}{3}
\Def{while_xvada_nt}{17}
\Def{while_xvada_T}{18}
\Def{while_xvada_A}{43}
\Def{while_xvada_IA}{2.65}
\Def{while_xvada_S}{114}
\Def{while_xvada_D}{5}
\Def{lisp_treevada_nt}{18}
\Def{lisp_treevada_T}{40}
\Def{lisp_treevada_A}{95}
\Def{lisp_treevada_IA}{1.65}
\Def{lisp_treevada_S}{157}
\Def{lisp_treevada_D}{5}
\Def{lisp_Kedavra_nt}{5}
\Def{lisp_Kedavra_T}{40}
\Def{lisp_Kedavra_A}{69}
\Def{lisp_Kedavra_IA}{1.16}
\Def{lisp_Kedavra_S}{80}
\Def{lisp_Kedavra_D}{3}
\Def{lisp_xvada_nt}{8}
\Def{lisp_xvada_T}{90}
\Def{lisp_xvada_A}{108}
\Def{lisp_xvada_IA}{1.31}
\Def{lisp_xvada_S}{142}
\Def{lisp_xvada_D}{4}
\Def{xml_treevada_nt}{9}
\Def{xml_treevada_T}{58}
\Def{xml_treevada_A}{78}
\Def{xml_treevada_IA}{2.00}
\Def{xml_treevada_S}{156}
\Def{xml_treevada_D}{4}
\Def{xml_Kedavra_nt}{8}
\Def{xml_Kedavra_T}{58}
\Def{xml_Kedavra_A}{122}
\Def{xml_Kedavra_IA}{1.52}
\Def{xml_Kedavra_S}{185}
\Def{xml_Kedavra_D}{4}
\Def{xml_xvada_nt}{12}
\Def{xml_xvada_T}{58}
\Def{xml_xvada_A}{79}
\Def{xml_xvada_IA}{1.84}
\Def{xml_xvada_S}{145}
\Def{xml_xvada_D}{5}
\Def{json_treevada_nt}{17}
\Def{json_treevada_T}{82}
\Def{json_treevada_A}{178}
\Def{json_treevada_IA}{1.17}
\Def{json_treevada_S}{208}
\Def{json_treevada_D}{5}
\Def{json_Kedavra_nt}{13}
\Def{json_Kedavra_T}{74}
\Def{json_Kedavra_A}{114}
\Def{json_Kedavra_IA}{1.20}
\Def{json_Kedavra_S}{137}
\Def{json_Kedavra_D}{4}
\Def{json_xvada_nt}{18}
\Def{json_xvada_T}{74}
\Def{json_xvada_A}{117}
\Def{json_xvada_IA}{1.35}
\Def{json_xvada_S}{158}
\Def{json_xvada_D}{5}
\Def{tinyc_treevada_nt}{30}
\Def{tinyc_treevada_T}{51}
\Def{tinyc_treevada_A}{150}
\Def{tinyc_treevada_IA}{1.76}
\Def{tinyc_treevada_S}{264}
\Def{tinyc_treevada_D}{5}
\Def{tinyc_Kedavra_nt}{10}
\Def{tinyc_Kedavra_T}{49}
\Def{tinyc_Kedavra_A}{57}
\Def{tinyc_Kedavra_IA}{1.56}
\Def{tinyc_Kedavra_S}{89}
\Def{tinyc_Kedavra_D}{5}
\Def{tinyc_xvada_nt}{28}
\Def{tinyc_xvada_T}{50}
\Def{tinyc_xvada_A}{138}
\Def{tinyc_xvada_IA}{1.88}
\Def{tinyc_xvada_S}{259}
\Def{tinyc_xvada_D}{5}
\Def{c-500_treevada_nt}{41}
\Def{c-500_treevada_T}{51}
\Def{c-500_treevada_A}{188}
\Def{c-500_treevada_IA}{1.84}
\Def{c-500_treevada_S}{345}
\Def{c-500_treevada_D}{5}
\Def{c-500_Kedavra_nt}{10}
\Def{c-500_Kedavra_T}{49}
\Def{c-500_Kedavra_A}{58}
\Def{c-500_Kedavra_IA}{1.59}
\Def{c-500_Kedavra_S}{92}
\Def{c-500_Kedavra_D}{6}
\Def{c-500_xvada_nt}{36}
\Def{c-500_xvada_T}{50}
\Def{c-500_xvada_A}{164}
\Def{c-500_xvada_IA}{1.93}
\Def{c-500_xvada_S}{316}
\Def{c-500_xvada_D}{5}
\Def{tiny_treevada_nt}{29}
\Def{tiny_treevada_T}{72}
\Def{tiny_treevada_A}{129}
\Def{tiny_treevada_IA}{1.46}
\Def{tiny_treevada_S}{188}
\Def{tiny_treevada_D}{6}
\Def{tiny_Kedavra_nt}{17}
\Def{tiny_Kedavra_T}{70}
\Def{tiny_Kedavra_A}{84}
\Def{tiny_Kedavra_IA}{1.20}
\Def{tiny_Kedavra_S}{101}
\Def{tiny_Kedavra_D}{7}
\Def{tiny_xvada_nt}{24}
\Def{tiny_xvada_T}{72}
\Def{tiny_xvada_A}{120}
\Def{tiny_xvada_IA}{1.48}
\Def{tiny_xvada_S}{177}
\Def{tiny_xvada_D}{6}
\Def{minic_treevada_nt}{39}
\Def{minic_treevada_T}{82}
\Def{minic_treevada_A}{209}
\Def{minic_treevada_IA}{1.80}
\Def{minic_treevada_S}{376}
\Def{minic_treevada_D}{6}
\Def{minic_Kedavra_nt}{19}
\Def{minic_Kedavra_T}{55}
\Def{minic_Kedavra_A}{75}
\Def{minic_Kedavra_IA}{1.69}
\Def{minic_Kedavra_S}{126}
\Def{minic_Kedavra_D}{8}
\Def{minic_xvada_nt}{42}
\Def{minic_xvada_T}{85}
\Def{minic_xvada_A}{176}
\Def{minic_xvada_IA}{1.65}
\Def{minic_xvada_S}{290}
\Def{minic_xvada_D}{7}
\Def{curl_treevada_nt}{18}
\Def{curl_treevada_T}{82}
\Def{curl_treevada_A}{135}
\Def{curl_treevada_IA}{1.38}
\Def{curl_treevada_S}{186}
\Def{curl_treevada_D}{4}
\Def{curl_Kedavra_nt}{23}
\Def{curl_Kedavra_T}{80}
\Def{curl_Kedavra_A}{124}
\Def{curl_Kedavra_IA}{1.17}
\Def{curl_Kedavra_S}{146}
\Def{curl_Kedavra_D}{6}
\Def{curl_xvada_nt}{22}
\Def{curl_xvada_T}{104}
\Def{curl_xvada_A}{208}
\Def{curl_xvada_IA}{1.30}
\Def{curl_xvada_S}{271}
\Def{curl_xvada_D}{5}
\Def{Avg_lo_treevada_nt}{24}
\Def{Avg_lo_treevada_T}{60}
\Def{Avg_lo_treevada_A}{131}
\Def{Avg_lo_treevada_IA}{1.64}
\Def{Avg_lo_treevada_S}{212}
\Def{Avg_lo_treevada_D}{5}
\Def{Avg_lo_Kedavra_nt}{12}
\Def{Avg_lo_Kedavra_T}{56}
\Def{Avg_lo_Kedavra_A}{81}
\Def{Avg_lo_Kedavra_IA}{1.6}
\Def{Avg_lo_Kedavra_S}{111}
\Def{Avg_lo_Kedavra_D}{5}
\Def{Avg_lo_xvada_nt}{22}
\Def{Avg_lo_xvada_T}{67}
\Def{Avg_lo_xvada_A}{123}
\Def{Avg_lo_xvada_IA}{1.66}
\Def{Avg_lo_xvada_S}{197}
\Def{Avg_lo_xvada_D}{5}
\Def{liquid_treevada_nt}{27}
\Def{liquid_treevada_T}{117}
\Def{liquid_treevada_A}{182}
\Def{liquid_treevada_IA}{4.74}
\Def{liquid_treevada_S}{863}
\Def{liquid_treevada_D}{4}
\Def{liquid_Kedavra_nt}{}
\Def{liquid_Kedavra_T}{}
\Def{liquid_Kedavra_A}{}
\Def{liquid_Kedavra_IA}{}
\Def{liquid_Kedavra_S}{}
\Def{liquid_Kedavra_D}{}
\Def{liquid_xvada_nt}{68}
\Def{liquid_xvada_T}{124}
\Def{liquid_xvada_A}{604}
\Def{liquid_xvada_IA}{1.63}
\Def{liquid_xvada_S}{984}
\Def{liquid_xvada_D}{4}
\Def{lua_treevada_nt}{55}
\Def{lua_treevada_T}{108}
\Def{lua_treevada_A}{310}
\Def{lua_treevada_IA}{4.71}
\Def{lua_treevada_S}{1,460}
\Def{lua_treevada_D}{4}
\Def{lua_Kedavra_nt}{}
\Def{lua_Kedavra_T}{}
\Def{lua_Kedavra_A}{}
\Def{lua_Kedavra_IA}{}
\Def{lua_Kedavra_S}{}
\Def{lua_Kedavra_D}{}
\Def{lua_xvada_nt}{110}
\Def{lua_xvada_T}{109}
\Def{lua_xvada_A}{526}
\Def{lua_xvada_IA}{2.65}
\Def{lua_xvada_S}{1,394}
\Def{lua_xvada_D}{5}
\Def{c_treevada_nt}{88}
\Def{c_treevada_T}{146}
\Def{c_treevada_A}{643}
\Def{c_treevada_IA}{7.44}
\Def{c_treevada_S}{4,786}
\Def{c_treevada_D}{6}
\Def{c_Kedavra_nt}{}
\Def{c_Kedavra_T}{}
\Def{c_Kedavra_A}{}
\Def{c_Kedavra_IA}{}
\Def{c_Kedavra_S}{}
\Def{c_Kedavra_D}{}
\Def{c_xvada_nt}{146}
\Def{c_xvada_T}{126}
\Def{c_xvada_A}{985}
\Def{c_xvada_IA}{2.92}
\Def{c_xvada_S}{2,880}
\Def{c_xvada_D}{7}
\Def{java_treevada_nt}{47}
\Def{java_treevada_T}{114}
\Def{java_treevada_A}{324}
\Def{java_treevada_IA}{5.79}
\Def{java_treevada_S}{1,876}
\Def{java_treevada_D}{4}
\Def{java_Kedavra_nt}{}
\Def{java_Kedavra_T}{}
\Def{java_Kedavra_A}{}
\Def{java_Kedavra_IA}{}
\Def{java_Kedavra_S}{}
\Def{java_Kedavra_D}{}
\Def{java_xvada_nt}{135}
\Def{java_xvada_T}{117}
\Def{java_xvada_A}{590}
\Def{java_xvada_IA}{2.18}
\Def{java_xvada_S}{1,287}
\Def{java_xvada_D}{6}
\Def{cpp_treevada_nt}{88}
\Def{cpp_treevada_T}{148}
\Def{cpp_treevada_A}{542}
\Def{cpp_treevada_IA}{9.40}
\Def{cpp_treevada_S}{5,093}
\Def{cpp_treevada_D}{5}
\Def{cpp_Kedavra_nt}{}
\Def{cpp_Kedavra_T}{}
\Def{cpp_Kedavra_A}{}
\Def{cpp_Kedavra_IA}{}
\Def{cpp_Kedavra_S}{}
\Def{cpp_Kedavra_D}{}
\Def{cpp_xvada_nt}{141}
\Def{cpp_xvada_T}{133}
\Def{cpp_xvada_A}{771}
\Def{cpp_xvada_IA}{4.89}
\Def{cpp_xvada_S}{3768}
\Def{cpp_xvada_D}{6}
\Def{rust_treevada_nt}{77}
\Def{rust_treevada_T}{180}
\Def{rust_treevada_A}{562}
\Def{rust_treevada_IA}{6.35}
\Def{rust_treevada_S}{3,569}
\Def{rust_treevada_D}{5}
\Def{rust_Kedavra_nt}{}
\Def{rust_Kedavra_T}{}
\Def{rust_Kedavra_A}{}
\Def{rust_Kedavra_IA}{}
\Def{rust_Kedavra_S}{}
\Def{rust_Kedavra_D}{}
\Def{rust_xvada_nt}{135}
\Def{rust_xvada_T}{143}
\Def{rust_xvada_A}{1,053}
\Def{rust_xvada_IA}{2.54}
\Def{rust_xvada_S}{2,673}
\Def{rust_xvada_D}{6}
\Def{mysql_treevada_nt}{70}
\Def{mysql_treevada_T}{265}
\Def{mysql_treevada_A}{405}
\Def{mysql_treevada_IA}{5.89}
\Def{mysql_treevada_S}{2,386}
\Def{mysql_treevada_D}{4}
\Def{mysql_Kedavra_nt}{}
\Def{mysql_Kedavra_T}{}
\Def{mysql_Kedavra_A}{}
\Def{mysql_Kedavra_IA}{}
\Def{mysql_Kedavra_S}{}
\Def{mysql_Kedavra_D}{}
\Def{mysql_xvada_nt}{130}
\Def{mysql_xvada_T}{378}
\Def{mysql_xvada_A}{798}
\Def{mysql_xvada_IA}{3.88}
\Def{mysql_xvada_S}{3,096}
\Def{mysql_xvada_D}{6}
\Def{Avg_hi_treevada_nt}{65}
\Def{Avg_hi_treevada_T}{154}
\Def{Avg_hi_treevada_A}{424}
\Def{Avg_hi_treevada_IA}{6}
\Def{Avg_hi_treevada_S}{2,862}
\Def{Avg_hi_treevada_D}{5}
\Def{Avg_hi_Kedavra_nt}{}
\Def{Avg_hi_Kedavra_T}{}
\Def{Avg_hi_Kedavra_A}{}
\Def{Avg_hi_Kedavra_IA}{}
\Def{Avg_hi_Kedavra_S}{}
\Def{Avg_hi_Kedavra_D}{}
\Def{Avg_hi_xvada_nt}{124}
\Def{Avg_hi_xvada_T}{161}
\Def{Avg_hi_xvada_A}{761}
\Def{Avg_hi_xvada_IA}{2.96}
\Def{Avg_hi_xvada_S}{2,297}
\Def{Avg_hi_xvada_D}{6}

\Def{turtle_xvada_precision_without_hdd}{.82}
\Def{turtle_xvada_recall_without_hdd}{1.00}
\Def{turtle_xvada_fscore_without_hdd}{.90}
\Def{turtle_xvada_precision_with_hdd}{.82}
\Def{turtle_xvada_recall_with_hdd}{1.00}
\Def{turtle_xvada_fscore_with_hdd}{.90}
\Def{turtle_xvada_hdd_time}{.9}
\Def{turtle_xvada_total_time}{15}
\Def{while_xvada_precision_without_hdd}{1.00}
\Def{while_xvada_recall_without_hdd}{.86}
\Def{while_xvada_fscore_without_hdd}{.92}
\Def{while_xvada_precision_with_hdd}{1.00}
\Def{while_xvada_recall_with_hdd}{.86}
\Def{while_xvada_fscore_with_hdd}{.92}
\Def{while_xvada_hdd_time}{.6}
\Def{while_xvada_total_time}{294}
\Def{lisp_xvada_precision_without_hdd}{.90}
\Def{lisp_xvada_recall_without_hdd}{.21}
\Def{lisp_xvada_fscore_without_hdd}{.34}
\Def{lisp_xvada_precision_with_hdd}{.91}
\Def{lisp_xvada_recall_with_hdd}{.26}
\Def{lisp_xvada_fscore_with_hdd}{.40}
\Def{lisp_xvada_hdd_time}{1.1}
\Def{lisp_xvada_total_time}{52}
\Def{xml_xvada_precision_without_hdd}{1.00}
\Def{xml_xvada_recall_without_hdd}{.27}
\Def{xml_xvada_fscore_without_hdd}{.42}
\Def{xml_xvada_precision_with_hdd}{1.00}
\Def{xml_xvada_recall_with_hdd}{.27}
\Def{xml_xvada_fscore_with_hdd}{.42}
\Def{xml_xvada_hdd_time}{.6}
\Def{xml_xvada_total_time}{117}
\Def{json_xvada_precision_without_hdd}{.98}
\Def{json_xvada_recall_without_hdd}{.48}
\Def{json_xvada_fscore_without_hdd}{.64}
\Def{json_xvada_precision_with_hdd}{.99}
\Def{json_xvada_recall_with_hdd}{.60}
\Def{json_xvada_fscore_with_hdd}{.75}
\Def{json_xvada_hdd_time}{.1}
\Def{json_xvada_total_time}{4}
\Def{tinyc_xvada_precision_without_hdd}{.92}
\Def{tinyc_xvada_recall_without_hdd}{.06}
\Def{tinyc_xvada_fscore_without_hdd}{.12}
\Def{tinyc_xvada_precision_with_hdd}{.93}
\Def{tinyc_xvada_recall_with_hdd}{.09}
\Def{tinyc_xvada_fscore_with_hdd}{.17}
\Def{tinyc_xvada_hdd_time}{.4}
\Def{tinyc_xvada_total_time}{92}
\Def{c-500_xvada_precision_without_hdd}{.30}
\Def{c-500_xvada_recall_without_hdd}{.67}
\Def{c-500_xvada_fscore_without_hdd}{.42}
\Def{c-500_xvada_precision_with_hdd}{.35}
\Def{c-500_xvada_recall_with_hdd}{.81}
\Def{c-500_xvada_fscore_with_hdd}{.48}
\Def{c-500_xvada_hdd_time}{9.3}
\Def{c-500_xvada_total_time}{861}
\Def{tiny_xvada_precision_without_hdd}{.96}
\Def{tiny_xvada_recall_without_hdd}{.41}
\Def{tiny_xvada_fscore_without_hdd}{.57}
\Def{tiny_xvada_precision_with_hdd}{.95}
\Def{tiny_xvada_recall_with_hdd}{.41}
\Def{tiny_xvada_fscore_with_hdd}{.57}
\Def{tiny_xvada_hdd_time}{.1}
\Def{tiny_xvada_total_time}{22}
\Def{minic_xvada_precision_without_hdd}{.88}
\Def{minic_xvada_recall_without_hdd}{.58}
\Def{minic_xvada_fscore_without_hdd}{.70}
\Def{minic_xvada_precision_with_hdd}{.87}
\Def{minic_xvada_recall_with_hdd}{.58}
\Def{minic_xvada_fscore_with_hdd}{.70}
\Def{minic_xvada_hdd_time}{.9}
\Def{minic_xvada_total_time}{537}
\Def{curl_xvada_precision_without_hdd}{.70}
\Def{curl_xvada_recall_without_hdd}{.33}
\Def{curl_xvada_fscore_without_hdd}{.45}
\Def{curl_xvada_precision_with_hdd}{.72}
\Def{curl_xvada_recall_with_hdd}{.35}
\Def{curl_xvada_fscore_with_hdd}{.47}
\Def{curl_xvada_hdd_time}{.3}
\Def{curl_xvada_total_time}{67}
\Def{Avg_lo_xvada_precision_without_hdd}{.85}
\Def{Avg_lo_xvada_recall_without_hdd}{.49}
\Def{Avg_lo_xvada_fscore_without_hdd}{.55}
\Def{Avg_lo_xvada_precision_with_hdd}{.85}
\Def{Avg_lo_xvada_recall_with_hdd}{.52}
\Def{Avg_lo_xvada_fscore_with_hdd}{.58}
\Def{Avg_lo_xvada_hdd_time}{1.4}
\Def{Avg_lo_xvada_total_time}{206}
\Def{liquid_xvada_precision_without_hdd}{.76}
\Def{liquid_xvada_recall_without_hdd}{.29}
\Def{liquid_xvada_fscore_without_hdd}{.42}
\Def{liquid_xvada_precision_with_hdd}{.99}
\Def{liquid_xvada_recall_with_hdd}{.52}
\Def{liquid_xvada_fscore_with_hdd}{.68}
\Def{liquid_xvada_hdd_time}{4.1}
\Def{liquid_xvada_total_time}{160}
\Def{lua_xvada_precision_without_hdd}{.70}
\Def{lua_xvada_recall_without_hdd}{.01}
\Def{lua_xvada_fscore_without_hdd}{.02}
\Def{lua_xvada_precision_with_hdd}{.68}
\Def{lua_xvada_recall_with_hdd}{.05}
\Def{lua_xvada_fscore_with_hdd}{.09}
\Def{lua_xvada_hdd_time}{1.4}
\Def{lua_xvada_total_time}{225}
\Def{c_xvada_precision_without_hdd}{.20}
\Def{c_xvada_recall_without_hdd}{.55}
\Def{c_xvada_fscore_without_hdd}{.29}
\Def{c_xvada_precision_with_hdd}{.35}
\Def{c_xvada_recall_with_hdd}{.57}
\Def{c_xvada_fscore_with_hdd}{.43}
\Def{c_xvada_hdd_time}{7.5}
\Def{c_xvada_total_time}{2,798}
\Def{java_xvada_precision_without_hdd}{.64}
\Def{java_xvada_recall_without_hdd}{.27}
\Def{java_xvada_fscore_without_hdd}{.38}
\Def{java_xvada_precision_with_hdd}{.69}
\Def{java_xvada_recall_with_hdd}{.27}
\Def{java_xvada_fscore_with_hdd}{.39}
\Def{java_xvada_hdd_time}{4.1}
\Def{java_xvada_total_time}{3,066}
\Def{cpp_xvada_precision_without_hdd}{.33}
\Def{cpp_xvada_recall_without_hdd}{.09}
\Def{cpp_xvada_fscore_without_hdd}{.14}
\Def{cpp_xvada_precision_with_hdd}{.35}
\Def{cpp_xvada_recall_with_hdd}{.09}
\Def{cpp_xvada_fscore_with_hdd}{.14}
\Def{cpp_xvada_hdd_time}{76.5}
\Def{cpp_xvada_total_time}{34,099}
\Def{rust_xvada_precision_without_hdd}{.62}
\Def{rust_xvada_recall_without_hdd}{.53}
\Def{rust_xvada_fscore_without_hdd}{.57}
\Def{rust_xvada_precision_with_hdd}{.52}
\Def{rust_xvada_recall_with_hdd}{.54}
\Def{rust_xvada_fscore_with_hdd}{.53}
\Def{rust_xvada_hdd_time}{10.7}
\Def{rust_xvada_total_time}{9,548}
\Def{mysql_xvada_precision_without_hdd}{.61}
\Def{mysql_xvada_recall_without_hdd}{.00}
\Def{mysql_xvada_fscore_without_hdd}{.01}
\Def{mysql_xvada_precision_with_hdd}{.61}
\Def{mysql_xvada_recall_with_hdd}{.00}
\Def{mysql_xvada_fscore_with_hdd}{.01}
\Def{mysql_xvada_hdd_time}{35.5}
\Def{mysql_xvada_total_time}{5,488}
\Def{Avg_hi_xvada_precision_without_hdd}{.55}
\Def{Avg_hi_xvada_recall_without_hdd}{.25}
\Def{Avg_hi_xvada_fscore_without_hdd}{.26}
\Def{Avg_hi_xvada_precision_with_hdd}{.60}
\Def{Avg_hi_xvada_recall_with_hdd}{.29}
\Def{Avg_hi_xvada_fscore_with_hdd}{.32}
\Def{Avg_hi_xvada_hdd_time}{20.0}
\Def{Avg_hi_xvada_total_time}{7,912}

\begin{document}

\title{Toward Inferring Accurate Context-free Grammars for Big Languages in a Black-box Setting}

\author{
\IEEEauthorblockN{
Mohammad Rifat Arefin\IEEEauthorrefmark{1},
Nuhiat Arefin\IEEEauthorrefmark{2},
Shanto Rahman\IEEEauthorrefmark{3},
Christoph Csallner\IEEEauthorrefmark{1}
}
\IEEEauthorblockA{\IEEEauthorrefmark{1}
Dept. of Computer Science and Engineering,
University of Texas at Arlington,
USA\\
mxa7262@mavs.uta.edu, csallner@uta.edu}
\IEEEauthorblockA{\IEEEauthorrefmark{2}
Dept. of Computer Science and Engineering,
Islamic University of Technology,
Gazipur, Bangladesh\\
nuhiatarefin@iut-dhaka.edu}
\IEEEauthorblockA{\IEEEauthorrefmark{3}
Dept. of Computer Science,
University of Texas at Austin,
USA\\
shanto.rahman@utexas.edu}
}

\maketitle

\begin{abstract}
Black-box context-free grammar inference is crucial for program analysis, reverse engineering, program understanding, fuzzing, and security. However, existing approaches such as \arvada{}, \treevada{}, \kedavra{}, and \crucio{} struggle with scalability, accuracy, and grammar readability, especially on larger languages. To address this challenge, we introduce \Tool{} with several new techniques for deterministic inference of context-free grammars. In an empirical comparison that avoids several pitfalls of recent studies, \Tool{} improves on the highest-scoring competitor (\treevada{}) both in grammar accuracy and grammar compactness. \toolName{} also found a CVE in the widely used Python Liquid engine. Fuzzing based on the \toolName{}-inferred grammar found five more bugs, which the Python Liquid developers fixed based on our bug reports. \toolName{} and all experimental data and scripts are publicly available.
\end{abstract}


\npdecimalsign{.}

\section{Introduction}
Black-box inference of context-free grammars from sample programs faces two key challenges. First, existing approaches do not scale to larger languages, at best producing grammars of low accuracy. Second, inferred grammars are excessively large and cryptic, making inferred grammars difficult for humans to interpret and use.

These two challenges are important as, first, many languages of practical use are large and complex~\cite{Crepinsek2010}. Second, besides fuzzing~\cite{aschermann2019nautilus, hodovan2018grammarinator} and security analysis~\cite{sassaman2013security}, other key use-cases of grammar inference include reverse engineering~\cite{caballero2007polyglot}, documentation, and human understanding of rare and under-documented commercial languages~\cite{lammel2001semi}, so readability is also very important. At a high level, these challenges are important as context-free grammars are widely used in programming language design, documentation, and as an input for fuzzing systems that search for parser bugs~\cite{reghizzi2013formal, Aho2006, yang2011finding}.

Black-box grammar inference is a fundamentally hard problem~\cite{angluin1991won}, as we only have access to a black-box parser of an unknown language $\mathcal{L}$ and a few sample $\mathcal{L}$ programs. The sample programs may not cover all $\mathcal{L}$ language features and will almost certainly not cover all possible combinations of these language features.

Recent advances in black-box context-free grammar inference include \arvada~\cite{kulkarni2021learning}, \treevada~\cite{arefin2024fast}, \kedavra~\cite{li2024incremental}, and \crucio{}~\cite{li2026contextfreegrammarinferencecomplex}. The pioneering \arvada{} infers grammars directly from complete input strings while exploring multiple generalization paths. \arvada{} suffers from low accuracy, inefficient processing, and poor grammar readability. The more recent \treevada{}, \kedavra{}, and \crucio{} succeed at improving upon \arvada{}. However, these improvement still do not scale to large and complex languages though. They also only partially address \arvada{}'s grammar readability issues. Specifically, even if the inferred grammar only partially covers the target ``golden'' grammar, the inferred grammar has many more rules and relies on a cryptic naming scheme.

\Tool{} is a deterministic automated \treevada{}-style black-box grammar inference technique. \Tool{} contains several new techniques for iterative parse-tree construction, including treating each bracketed token sequence as a candidate grammar rule alternative, adding indirection to enable learning of non-recursive rules, and batch grammar generalization processing. The latter translates a successful grammar generalization into an update of the current ranked list of remaining candidate grammar generalization steps, to avoid the expensive re-ranking of all parse trees after each generalization step. \toolName{} further decomposes complex inferred parse trees via Hierarchical Delta Debugging (HDD)~\cite{hdd} to learn grammar rule alternatives and infers descriptive non-terminal names via LLM queries.

In an empirical comparison that addresses several experimental design issues in recent related studies (e.g., \gladeRep{}~\cite{bendrissou2022synthesizing} and \crucio{}), \treevada{} and \toolName{} were the only approaches that could successfully process all 17~languages of various sizes. On smaller languages \toolName{} yielded grammars of similar accuracy, while achieving much higher average accuracy on larger languages. At the same time, \toolName{}-inferred grammars were on average more compact. While running on Python Liquid, \toolName{} found a denial-of-service CVE. Using the \toolName{}-inferred grammar, an off-the-shelf fuzzer then found five bugs the Python Liquid developers fixed based on our bug reports.
To summarize, the paper makes the following major contributions.
\begin{itemize}

\item \toolName{} is the first black-box approach for inferring context-free grammars that scales to larger languages and produces grammars that are compact with high accuracy.

\item The paper identifies pitfalls in recent work that can inflate F1 scores, including incompatible precision/recall grammar interpretations and duplicate-heavy test suites.

\item In an empirical comparison on larger languages, \Tool{} improves on the highest-scoring competitor (\treevada{}) both in grammar accuracy and grammar compactness.

\item \toolName{} also found a CVE in the widely used Python Liquid engine. Fuzzing based on the \toolName{}-inferred grammar found five more since-fixed Python Liquid bugs.

\item The \Tool{} implementation, scripts, and data (including for replicating the experiments) are freely available online: \url{https://github.com/rifatarefin/xvada}

\end{itemize}

\section{Background}

Almost all programming languages employ context-free grammars in their design~\cite{reghizzi2013formal}. Similarly, many popular languages are at least partially documented via a context-free grammar. Context-free grammars are also a popular input format for fuzzers. Following is relevant context of the \arvada{} and \treevada{} approaches \toolName{} builds on, as well as issues we identified in the experimental design of \crucio{} we aim to improve upon in our evaluation.

\subsection{Grammar Inference via Parse Tree Generalization}

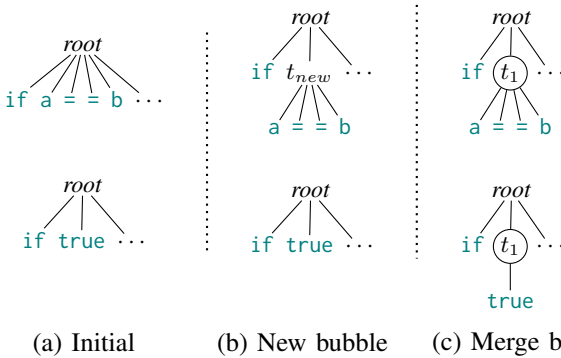
\begin{figure}[h!t]
\centering
\begin{tikzpicture}[>=Stealth, every node/.style={inner sep=1pt}]

\node (LeftTop) {
  \begin{forest}
    for tree={l sep=6pt, s sep=2pt, font=\small}
    [\textit{root}
      [\term{if}]
      [\term{a}]
      [{\term{=}}]
      [{\term{=}}]
      [\term{b}]
      [{\(\cdots\)}] %
    ]
  \end{forest}
};

\node (LeftBot) [below=.8cm of LeftTop] {
  \begin{forest}
    for tree={l sep=6pt, s sep=2pt, font=\small}
    [\textit{root}
      [\term{if}]
      [\term{true}]
      [{\(\cdots\)}]
    ]
  \end{forest}
};
\node (a) [below= of LeftBot.south] {(a) Initial};
\node (MidTop) [right= of LeftTop] {
  \begin{forest}
    for tree={l sep=6pt, s sep=2pt, font=\small}
    [\textit{root}
      [\term{if}]
      [$t_{new}$[\term{a}]
      [{\term{=}}]
      [{\term{=}}]
      [\term{b}]]
      [{\(\cdots\)}] %
    ]
  \end{forest}
};

\node (MidBot) [below=.5cm of MidTop] {
  \begin{forest}
    for tree={l sep=6pt, s sep=2pt, font=\small}
    [\textit{root}
      [\term{if}]
      [\term{true}]
      [{\(\cdots\)}]
    ]
  \end{forest}
};
\node (b) [below= of MidBot.south, right=1cm of a.east] {(b) New bubble};
\node (RightTop) [right= of MidTop] {
  \begin{forest}
    for tree={l sep=6pt, s sep=2pt, font=\small}
    [\textit{root}
      [\term{if}]
      [$t_1$, draw, circle[\term{a}]
      [{\term{=}}]
      [{\term{=}}]
      [\term{b}]]
      [{\(\cdots\)}] %
    ]
  \end{forest}
};

\node (RightBot) [below=.5cm of RightTop] {
  \begin{forest}
    for tree={l sep=6pt, s sep=2pt, font=\small}
    [\textit{root}
      [\term{if}]
      [$t_1$, draw, circle[\term{true}]]
      [{\(\cdots\)}]
    ]
  \end{forest}
};
\node (c) [below= of RightBot.south, right=.5cm of b.east] {(c) Merge bubble};
\coordinate (Sep) at ($(LeftTop.east)!0.5!(MidTop.west)$);

\draw[dotted, thick]
  (Sep |- LeftTop.north)
  --
  (Sep |- LeftBot.south);

\coordinate (SepR) at ($(MidTop.east)!0.5!(RightTop.west)$);

\draw[dotted, thick]
  (SepR |- MidTop.north)
  --
  (SepR |- MidBot.south);
\end{tikzpicture}
\caption{Example incremental parse tree construction: (a) initial flat parse trees, (b) node sequence \term{a==b} bubbled-up under new non-terminal $t_{new}$, (c) $t_{new}$ and \term{true} merged as $t_1$.}
\label{fig:bubble}
\end{figure}

\arvada{} introduced parse tree generalization via \textit{bubble-and-merge} for learning context-free grammars~\cite{kulkarni2021learning}. \arvada{} starts from initial flat trees, e.g., the two Figure~\ref{fig:bubble}a example trees. Figure~\ref{fig:bubble}b bubbles up the sibling node sequence \term{a==b} under the new non-terminal node $t_{new}$. \arvada{} then checks if it can replace $t_{new}$ with any existing non-terminal node and vice-versa.

As replacing (the implicit parent of) \term{true} with $t_{new}$ and vice versa here yields syntactically correct programs, \arvada{} treats \term{a==b} and \term{true} as alternatives of the same grammar rule. We thus merge $t_{new}$ and (the parent of) \term{true} and relabel them as $t_1$ in Figure~\ref{fig:bubble}c. The tree-induced grammar thus now includes the new rule $t_1 \to \term{a==b}\; | \;\term{true}$.

Instead of flat trees, \arvada{}'s successor \treevada{}~\cite{arefin2024fast} first pre-structures trees, based on the common structural assumption~\cite{sakakibara1992efficient} that brackets indicate concept nesting~\cite{van2019lightweight}.

Before accepting a candidate bubble into the parse trees, \arvada{}/\treevada{} heuristically validate it  via \textsc{CheckBubble}. \textsc{CheckBubble} samples (at most 50) candidate strings from the parse trees, by instantiating the candidate alternatives throughout the parse trees. If the black-box parser accepts all thereby generated programs, \textsc{CheckBubble} returns true and the new generalization is accepted. The limited sampling (up to 50) may miss invalid program instantiations and may thus merge invalid generalizations into the trees.

\subsection{Reassessing \crucio's Evaluation} 

The recent \crucio{} black-box grammar inference work reported high F1 scores for complex programming languages~\cite{li2026contextfreegrammarinferencecomplex}. Upon close examination, we find \textit{\textbf{two major issues}} with that study's evaluation design that make it hard to generalize these early findings.

\MyPara{Different Grammars for Precision and Recall Metrics}
A key limitation of \crucio{}'s evaluation methodology is that precision and recall are computed using different interpretations of the inferred grammar. Specifically, precision is measured using a grammar that instantiates each lexical class only with the concrete character sequences observed during training, whereas recall is measured using a different grammar that maps the same lexical class to a deterministic finite automaton (DFA), potentially accepting a much larger set of character sequences. As a result, the reported F1 score combines precision and recall computed on different grammars, making it difficult to interpret and compare fairly with approaches that evaluate both metrics using the same inferred grammar.

More formally, let $\tau$ denote an inferred lexical class, $S_\tau$ denote the set of tokens of type $\tau$ observed in the seed programs, and $D_\tau$ denote the language accepted by the DFA inferred for $\tau$, where typically $S_\tau \neq D_\tau$. During precision evaluation, \crucio{} generates programs by instantiating $\tau$ exclusively with tokens from $S_\tau$. During recall evaluation, however, an unseen token $x \notin S_\tau$ is considered an instance of $\tau$ whenever $x \in D_\tau$. Precision thus evaluates a grammar whose lexical productions are restricted to $S_\tau$, yet recall evaluates a more permissive grammar whose lexical productions are defined by $D_\tau$. For example, if the seed inputs contain only the identifier \term{FooBar}, precision is computed using a grammar that generates only \term{FooBar} for that identifier class, while recall is computed using a grammar whose inferred DFA may accept a much broader set of identifiers.

\MyPara{Precision Metric Does Not Sample Complete Programs} 
All recent grammar inference approaches evaluate precision by sampling programs from the inferred grammar. Earlier studies, including \arvada{}~\cite{kulkarni2021learning}, \treevada{}~\cite{arefin2024fast}, \kedavra{}~\cite{li2024incremental}  recursively generate an entire program by expanding grammar rules from the start symbol to the lexical terminals. We refer to this conventional strategy as \emph{full-program sampling} ($P_{sample}$). In contrast, \crucio{} introduces a mutation-based precision metric, denoted as $P_{swap}$. Rather than generating a complete program from scratch, $P_{swap}$ begins with an existing valid program accepted by the inferred grammar and randomly selects a non-terminal subtree to regenerate while keeping the remainder of the program unchanged.

\begin{figure}[h!t]
\centering

\centering

\begin{minipage}{0.48\columnwidth}
\centering
\resizebox{\linewidth}{!}{%
\begin{forest}
  base parse tree,
  tikz+={
    \begin{scope}[on background layer]
      \node[
        draw=blue!65!black,
        fill=blue!6,
        line width=.8pt,
        inner sep=0pt,
        rounded corners=0pt,
        fit=(expA)(eidA)(opA)(numA)(yA)(plusA)(twoA)
      ] {};
    \end{scope}
  }
  [stmt
    [id, ctx
      [\term{x}, ctx, font=\footnotesize]
    ]
    [assign, ctx
      [{\term{=}}, ctx, font=\footnotesize]
    ]
    [exp, blue hi, name=expA, before computing xy={s+=1.1mm}
      [id, blue hi, name=eidA
        [\term{y}, blue hi, name=yA, font=\footnotesize]
      ]
      [plus, blue hi, name=opA
        [\term{+}, blue hi, name=plusA, font=\footnotesize]
      ]
      [num, blue hi, name=numA
        [\term{2}, blue hi, name=twoA, font=\footnotesize]
      ]
    ]
    [semi, ctx, before computing xy={s+=1.1mm}
      [\term{;}, ctx, font=\footnotesize]
    ]
  ]
\end{forest}

}
\end{minipage}
\begin{minipage}{0.48\columnwidth}
\centering
\resizebox{\linewidth}{!}{%
\begin{forest}
  base parse tree,
  tikz+={
    \begin{scope}[on background layer]
      \node[
        draw=green!50!black,
        fill=green!7,
        line width=.8pt,
        inner sep=0pt,
        rounded corners=0pt,
        fit=(expB)(eidB)(opB)(numB)(yB)(timesB)(twoB)
      ] {};
    \end{scope}
  }
  [stmt
    [id, ctx
      [\term{x}, ctx, font=\footnotesize]
    ]
    [assign, ctx
      [{\term{=}}, ctx, font=\footnotesize]
    ]
    [exp, green hi, name=expB, before computing xy={s+=1.1mm}
      [id, green hi, name=eidB
        [\term{y}, green hi, name=yB, font=\footnotesize]
      ]
      [times, green hi, name=opB
        [\term{*}, green hi, name=timesB, font=\footnotesize]
      ]
      [num, green hi, name=numB
        [\term{2}, green hi, name=twoB, font=\footnotesize]
      ]
    ]
    [semi, ctx, before computing xy={s+=1.1mm}
      [\term{;}, ctx, font=\footnotesize]
    ]
  ]
\end{forest}
}
\end{minipage}

\caption{$P_{swap}$ mutates $exp$ while freezing $exp$'s context.}
\label{fig:side-by-side-derivations}

\end{figure}
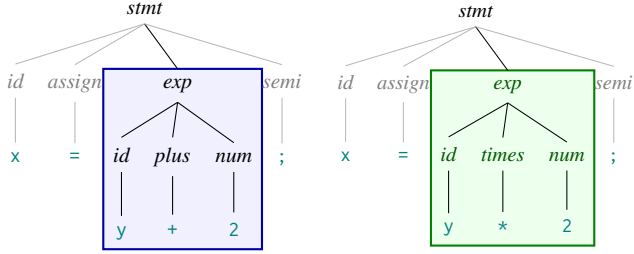

Figure~\ref{fig:side-by-side-derivations} illustrates this process. $P_{swap}$ randomly selects non-terminal \emph{exp}, freezes its left and right context, and only expands \emph{exp} to generate a concrete string fragment (\term{y*2}), which it then plugs back into the original program.

\forestset{
  base parse tree/.style={
    for tree={
      font=\itshape\footnotesize,
      l sep=6pt, 
      s sep=2pt,
      edge={-},
      parent anchor=south,
      child anchor=north,
    },
  },
  ctx/.style={
    text=gray,
    edge={gray!55},
  },
  blue hi/.style={
    text=black,
    line width=.8pt,
    edge={black},
  },
  green hi/.style={
    text=green!35!black,
    line width=.8pt,
    edge={black},
  },
}

Evaluating a grammar's generative capacity recursively from the top down (as in $P_{sample}$) exposes over-generalization. Lacking context-sensitivity, a flawed (context-free) grammar may incorrectly allow constructs such as a $while$ loop inside a variable assignment. $P_{swap}$ conceals such structural flaws. By freezing the mutation's context in the original valid program, $P_{swap}$ ensures that the program remains correct outside the (local) mutation, which artificially constrains the search space for generating invalid programs.

\begin{table}[h!t]
\centering
\caption{$P_{\text{sample}}$ vs $P_{\text{swap}}$ of \crucio{}'s artifact (excerpt).}
\label{table:p_swap}
\begin{tabular}{l|r|r|r|r}
\toprule
\multicolumn{1}{c|}{} & \multicolumn{1}{c|}{P-sample} & \multicolumn{1}{c|}{P-swap} & \multicolumn{1}{c|}{F1-sample} & \multicolumn{1}{c}{F1-swap} \\
\midrule
 \Use{lua_programname}  & \Use{lua_p_sample_and_std} & \Use{lua_p_swap_and_std} & \Use{lua_f1_sample_and_std} & \Use{lua_f1_swap_and_std}\\
 \Use{c_programname}  & \Use{c_p_sample_and_std} & \Use{c_p_swap_and_std} & \Use{c_f1_sample_and_std} & \Use{c_f1_swap_and_std}\\
 \Use{java_programname}  & \Use{java_p_sample_and_std} & \Use{java_p_swap_and_std} & \Use{java_f1_sample_and_std} & \Use{java_f1_swap_and_std}\\
 \Use{cpp_programname}  & \Use{cpp_p_sample_and_std} & \Use{cpp_p_swap_and_std} & \Use{cpp_f1_sample_and_std} & \Use{cpp_f1_swap_and_std}\\
 \Use{rust_programname}  & \Use{rust_p_sample_and_std} & \Use{rust_p_swap_and_std} & \Use{rust_f1_sample_and_std} & \Use{rust_f1_swap_and_std}\\
\bottomrule
\end{tabular}
\end{table}

Based on the \crucio{} artifact logs, Table~\ref{table:p_swap} shows the impact of using $P_{swap}$ or $P_{sample}$ in \crucio{} for five languages used in previous studies. $P_{swap}$ is often an order of magnitude higher, also increasing the \fscore{}.

$P_{swap}$ also diverges from inferred grammars' practical utility. The primary downstream consumers of these tools, such as syntax-directed fuzzing engines~\cite{hodovan2018grammarinator, fandango2025amaya} typically operate by expanding a grammar continuously from the start rule to generate novel inputs. A precision metric based on $P_{swap}$ effectively evaluates the stability of the grammar under single-point mutation rather than its standalone generative robustness. While mutation-based generation is a valid testing strategy, using it to define the precision of the grammar itself misrepresents the artifact's reliability when deployed in standard, end-to-end generative tasks.

\section{\Tool{}}
\label{sec:methodology}

We present \toolName{}, a deterministic black-box CFG inference tool that iteratively builds parse trees by merging tree non-terminals. This section describes \Tool{}'s building blocks in pipeline order. After inference, \toolName{} optionally queries an LLM to label each non-terminal with a summary of the strings it derives. For readability, we use these LLM-generated non-terminal names also in the intermediate \toolName{} example trees throughout this section.

\subsection{Initial Parse-trees}

\subsubsection{Tokenization}

Before constructing parse trees, \toolName{} tokenizes the input programs.
Each whitespace, special, or non-ASCII character \toolName{} treats as an individual token. Each contiguous letter sequence and each contiguous digit sequence \toolName{} groups into a token labeled by the grouped sequence. For example, \toolName{} maps \term{\$\$FooBar123} to four tokens: \term{\$}, \term{\$}, \term{FooBar}, and \term{123}.

\subsubsection{Minimizing whitespace}

To reduce the size of an input program's token sequence and the subsequent search space, we iteratively trim whitespace. This process ensures that the input program remains valid, i.e., it retains required whitespace. \toolName{} scans the token sequence left to right. Whenever we encounter a whitespace, we temporarily remove it and query the oracle on the resulting character sequence. If the oracle accepts the input, we omit the whitespace; otherwise, we restore it. For example, with a C-language compiler \toolName{} trims \term{int\textvisiblespace{}\textvisiblespace{}a\textvisiblespace{}=\textvisiblespace{}\textvisiblespace{}1\textvisiblespace{};} to \term{int\textvisiblespace{}a=1;}.

\subsubsection{String-literal guardrail}

Prior approaches (including \treevada{}, \kedavra{}, and Crucio) commonly treat a sequence wrapped by single or double quotes such as \term{'a'} or \term{"big dog"} as a single string literal token. This heuristic works well for many languages, but it may be unsound when quotation marks serve other purposes. 

To avoid incorrectly mapping such sequences to a single token, we introduce an oracle-guided string-identification guardrail. When \toolName{} encounters a sequence enclosed by matching quotes or backticks (\term{'}, \term{"}, or \term{`}), we replace the enclosed contents with arbitrary characters and query the oracle. If the oracle accepts the modified program, \toolName{} treats this as evidence that the quoted contents are unconstrained string data and collapses the quote-enclosed sequence to a single token, so \term{"big dog"} would yield the three tokens \term{"}, \term{big dog}, and \term{"}. Otherwise, we keep the sequence's default tokenization.

\subsubsection{Bracket-based parse-tree initialization}

After tokenization, \toolName{} constructs for each seed program an initial flat parse tree. Following \treevada{}, \toolName{} is built on the ``soft'' assumption that many languages use \term{[]()\{\}} brackets as (recursive) nesting constructs. We thus use a bracket-based heuristic that moves each bracketed token sequence under a newly created child node, thereby introducing an initial syntactic hierarchy. Once this pre-structuring step is complete, we invoke \textsc{MergeAllValid}, which systematically merges any pair of non-terminal nodes it can interchange while preserving oracle-validity of the generated strings.

\subsection{Bracket-guided Bubble Selection} 
\label{sec:structure_infer}

\forestset{
  stmt tree/.style={
    for tree={
      l sep=2pt,
      s sep=1pt,
      font=\small,
      edge={-},
      child anchor=north,
    },
  }
}

\begin{figure}[t]
\centering

\begin{minipage}{0.5\columnwidth}
\centering
\resizebox{\linewidth}{!}{%
\begin{forest}
stmt tree,
tikz+={
  \begin{scope}[on background layer]
    \node[
      fill=blue!10,
      line width=.8pt,
      inner sep=0pt,
      rounded corners=0pt,
      fit=(h1)(h2)(h3)
    ] {};
  \end{scope}
}
[\textit{stmt}
  [\term{if}]
  [\textit{t\textsubscript{1}}
    [\term{(}]
    [\term{1}]
    [\term{+}]
    [\term{1}]
    [\term{)}]
  ]
  [\term{then}]
  [\term{x}]
  [{\term{=}}, name=h1]
  [\term{1}, name=h2]
  [\term{+}, name=h3]
  [\term{1}]
  [\term{end}]
]
\end{forest}
}
\end{minipage}
\begin{minipage}{0.45\columnwidth}
\centering
\resizebox{\linewidth}{!}{%
\begin{forest}
stmt tree,
tikz+={
  \begin{scope}[on background layer]
    \node[
      fill=blue!10,
      line width=.8pt,
      inner sep=0pt,
      rounded corners=0pt %
    ] {};
  \end{scope}
}
[\textit{stmt}
  [\term{if}]
  [\textit{t\textsubscript{1}}
    [\term{(}]
    [\term{1}]
    [\term{+}]
    [\term{1}]
    [\term{)}]
  ]
  [\term{then}]
  [\term{x}]
  [\textit{t\textsubscript{2}}, name=exp, draw, circle
  [\textit{t\textsubscript{2}}, name=h1, draw, circle [\term{=}, name=h4] ]
    [\term{1}, name=h2]
    [\term{+}, name=h3]
  ]
    [\term{1}, before computing xy={s+=2mm}]
  [\term{end}]
]
\end{forest}
}
\end{minipage}

\begin{minipage}{0.5\columnwidth}
\centering
\resizebox{\linewidth}{!}{%
\begin{forest}
stmt tree,
tikz+={
  \begin{scope}[on background layer]
    \node[
      fill=blue!10,
      line width=.8pt,
      inner sep=3pt,
      rounded corners=0pt,
      fit=(h1)(hplus)(h2)
    ] {};
  \end{scope}
}
[\textit{stmt}
  [\term{if}]
  [\textit{cnd}
    [\term{(}]
    [\term{1}, name=h1, draw=none]
    [\term{+}, name=hplus, draw=none]
    [\term{1}, name=h2, draw=none]
    [\term{)}]
  ]
  [\term{then}]
  [\term{x}]
  [{\term{=}}]
  [\term{1}]
  [\term{+}]
  [\term{1}]
  [\term{end}]
]
\end{forest}
}
\end{minipage}
\begin{minipage}{0.45\columnwidth}
\centering
\resizebox{\linewidth}{!}{%
\begin{forest}
stmt tree,
[\textit{stmt}
  [\term{if}]
  [\textit{cnd}
    [\term{(}]
    [\textit{exp}, name=e1
    [\textit{exp}[\term{1}, name=e2]]
    [\term{+}, name=e3]
    [\textit{exp}[\term{1}], name=e4]]
    [\term{)}]
  ]
  [\term{then}]
  [\term{x}]
  [{\term{=}}]
  [\textit{exp}, name=exp %
    [\textit{exp}[\term{1}, name=h1]]
    [\term{+}, name=hplus]
    [\textit{exp}[\term{1}, name=h2]]
  ]
  [\term{end}]
]
\end{forest}
}
\end{minipage}

\caption{Top: After ranking 30 candidates, \treevada{} merges the \term{=1+} bubble. Bottom: \toolName{} merges the bracketed \term{1+1}.
}
\label{fig:bracket_bubble}
\end{figure}
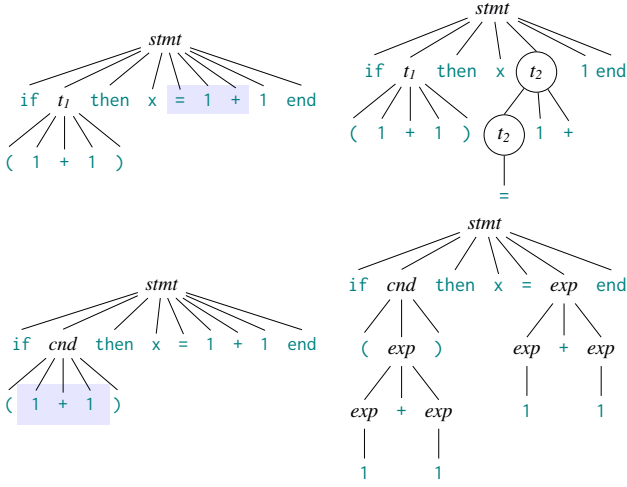

\begin{figure*}[!htbp]
\centering
\forestset{
  parse tree/.style={
    for tree={
      font=\itshape\small,
      inner sep=2pt,
      l sep=4pt,
      s sep=1pt,
      edge={-},
      child anchor=north,
      if n children=0{
        draw=none,
        font=\small,
        rounded corners=0pt,
        minimum width=0pt,
        minimum height=0pt,
      }{},
    },
  },
  ctx/.style={
    draw=gray!55,
    text=gray!65,
    edge={gray!55},
  },
  focus/.style={
    draw=blue!70!black,
    text=black,
    line width=.8pt,
    edge={black},
  },
  relabeled/.style={
    draw=green!50!black,
    text=green!35!black,
    line width=.8pt,
    edge={black},
  },
}

\begin{tikzpicture}[
  >=Stealth,
  tree holder/.style={inner sep=0pt},
  decision box/.style={
  draw,
  rounded corners=2pt,
  align=left,
  font=\small,
  inner sep=6pt,
  text width=3.8cm,
  fill=gray!5
},
  arrow label/.style={
    font=\small,
    align=center,
    fill=white,
    inner sep=1pt
  }
]

\node[tree holder] (T1) {
\begin{forest}
  parse tree
  [{\(\cdots\)}, ctx
    [{\(\cdots\)}, ctx]
    [expr
      [\term{-}]
      [num [\term{9}]]
      [\term{+}]
      [num [\term{5}]]
    ]
  ]
\end{forest}
};

\node[tree holder, right=1.5cm of T1] (T2) {
\begin{forest}
  parse tree
  [{\(\cdots\)}, ctx
    [{\(\cdots\)}, ctx]
    [expr
      [$t$, focus, circle
        [\term{-}]
        [num [\term{9}]]
      ]
      [\term{+}]
      [num [\term{5}]]
    ]
  ]
\end{forest}
};

\node[tree holder, right=3.5cm of T2] (T3) {
\begin{forest}
  parse tree
  [{\(\cdots\)}, ctx
    [{\(\cdots\)}, ctx]
    [expr
      [$t$, focus, circle
        [\term{-}]
        [num, [\term{9}]]
      ]
      [\term{+}]
      [int, focus, circle [num[\term{5}]]]
    ]
  ]
\end{forest}
};

\node[tree holder, right=3.6cm of T3] (T4) {
\begin{forest}
  parse tree
  [{\(\cdots\)}, ctx
    [{\(\cdots\)}, ctx]
    [expr
      [int, relabeled, circle
        [\term{-}]
        [num [\term{9}]]
      ]
      [\term{+}]
      [int, focus, circle [num[\term{5}]]]
    ]
  ]
\end{forest}
};

\coordinate (A23) at ($(T2.east)!0.5!(T3.west)$);
\coordinate (A34) at ($(T3.east)!0.5!(T4.west)$);

\node[decision box] (D2) at ($(A23)+(-.3cm,1.8cm)$) {
\textit{num} replacing $t$: \term{5 + 5}
{\color{green!60!black}\checkmark}\\
$t$ replacing \textit{num}: \term{--9 + -9}
{\color{red!70!black}\(\times\)}
};

\node[decision box] (D3) at ($(A34)+(-.3,1.8cm)$) {
\textit{int} replacing $t$: \term{5 + 5}
{\color{green!60!black}\checkmark}\\
$t$ replacing \textit{int}: \term{-9 + -9}
{\color{green!60!black}\checkmark}
};

\draw[->, thick] (T1.east) -- (T2.west)
  node[midway, above, arrow label] {Bubble\\$[$\term{-}$,\;$\emph{num}$]$};

\draw[->, thick] (T2.east) -- (T3.west)
  node[midway, below, arrow label] {Add indirection};

\draw[->, thick] (T3.east) -- (T4.west)
  node[midway, below, arrow label] {Merge};

\draw[->, dashed]
  ([xshift=-.5cm, yshift=.3cm]T2.east) -- (T2.east |- D2.south)
  node[midway, right, arrow label] {Sample strings};

\draw[->, dashed]
  ([xshift=-.5cm, yshift=.3cm]T3.east) -- (T3.east |- D3.south)
  node[midway, right, arrow label] {Sample strings};

\end{tikzpicture}
\caption{Non-recursive merge example, for which \arvada{} \& \treevada{} do not learn \emph{tiny}'s non-recursive int negation rule.
}
\label{fig:half-merge}

\end{figure*}
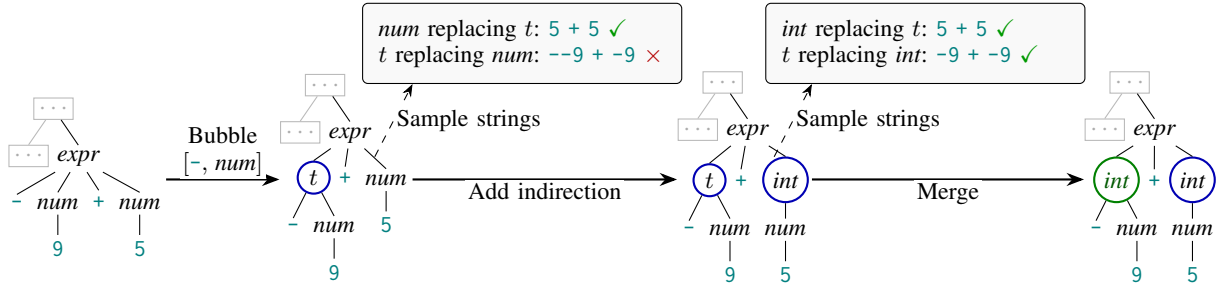

While \treevada{} pre-structures its parse trees along \term{[]()\{\}} brackets as (recursive) nesting constructs, \treevada{} does not exploit the bracketed sequence itself. As such a bracketed sequence can be a syntactic unit, \toolName{} treats each bracketed sequence as a candidate bubble, e.g., the \term{1+1} sequence in \term{()}-brackets the Figure~\ref{fig:bracket_bubble} Lua program. As with any candidate bubble \textit{b}, \toolName{} tries to merge all occurrences of \textit{b} in the seed programs in the same step, and succeeds here with merging both \term{1+1} instances with \term{1} into \textit{exp}.

While earlier approaches such as \arvada{} and \treevada{} may eventually also propose the same \term{1+1} bubble, these earlier approaches would do so as a result of their regular (expensive) bubble ranking process. Besides extra runtime, earlier tools may also first get confused by one of their heuristics, get stuck in their exploration, and thus never apply such a within-brackets bubble candidate. We thus prioritize bracket-derived bubble candidates over regular ones, and rank the resulting in-bracket bubble list using \treevada{}'s depth and length-aware heuristic.

\subsection{Indirection-based Merge to Learn Non-recursive Rules}

\arvada{}'s bidirectional replacement check prevents both \arvada{} and \treevada{} from learning certain non-recursive rules. Specifically, requiring bidirectional interchangeability of a candidate bubble with its direct or transitive child creates recursive grammar rules. Besides such recursive rules, \toolName{} also supports learning non-recursive rules via its indirection-based merge technique. The technique applies when unidirectional replacement succeeds, but the merge target non-terminal also occurs inside the candidate bubble. 

Such an in-bubble occurrence creates a recursive rule and thus a recursive replacement check, causing an otherwise valid merge to be rejected. To resolve this, we add a level of indirection to the target node. If the bidirectional check succeeds with this indirection we merge the bubbled-up node with the added node. If the merge-attempt fails we discard the changes. The tiny language's expression rules are an example. Specifically, tiny permits at most one negation per int:

\[
\begin{array}{l}
\textit{expr} \to \textit{expr op elem} \mid \textit{elem}\\
\textit{op} \to \term{+}\mid\term{-}\\
\textit{elem} \to \textit{ident} \mid \textit{int}\\
\textit{int} \to \textit{num} \mid \term{-}~\textit{num} \\
\textit{num} \to (\term{0} \mid \cdots \mid \term{9})+
\end{array}
\]

Figure~\ref{fig:half-merge} (left) illustrates this limitation for the subtree rooted at \textit{expr} in a seed program that contains the fragment \term{-9 + 5}. At this point (potentially among others) we consider \term{9} and \term{5} as equivalent at \emph{num}, e.g., after generating programs with fragments \term{-9 + 9} and \term{-5 + 5} the oracle accepted as valid, implying the rule
$\textit{num} \to \term{5} \mid \term{9}$.

\arvada{}/\treevada{} next consider candidate bubble $[$\term{-},~\emph{num}$]$ under a fresh non-terminal node $t$. Specifically, \textsc{CheckBubble} tests if $[$\term{-},~\emph{num}$]$ at $t$ is equivalent with the already established alternative expansions of an existing non-terminal in the tree such as \textit{num}.  If successful, this would merge $[$\term{-},~\emph{num}$]$ at $t$ with \term{5} and \term{9} at $num$, yielding the rule
$\textit{num} \to \term{-}~\textit{num} \mid \term{5} \mid \term{9}$.

In the equivalence test (via mutual replacement), \textit{num} replacing $t$ yields valid tiny programs with \emph{expr} such as \term{5~+~5}. But the reverse ($t$ replacing \textit{num}) produces invalid tiny programs, e.g., containing a \emph{expr} of \term{--9 + -9}. This triggers \arvada{} and \treevada{} to discard the $[$\term{-},~\emph{num}$]$ candidate bubble, which (for a seed program as in Figure~\ref{fig:half-merge}) makes them miss tiny's non-recursive integer negation rule. Instead they are left with the overly specific rule 
$\textit{expr} \to \term{-}~\textit{num}~\term{+}~\textit{num}$.

The key observation is that we can still generalize such an overly specific grammar rule, even in a non-recursive case as tiny's int negation. At the core, we want to test the equivalence of $[$\term{-},~\emph{num}$]$ with the existing derivations (\term{5} and \term{9}) of \textit{num}. Instead of merging \textit{t} with \textit{num} directly, we thus add a level of indirection, i.e., a new intermediate node between \textit{num} and its parent everywhere in our parse trees other than our candidate bubble \textit{t}. This yields the same equivalence test, but breaks recursion, is accepted by the oracle parser, and infers tiny's non-recursive int negation rule
$\textit{int} \to \term{-}~\textit{num}\mid\textit{num}$.

\subsection{Batch Bubble Processing}

\begin{figure*}[htbp]
\centering
\begin{tikzpicture}[
  >=Stealth,
  every node/.style={inner sep=1pt},
  flow/.style={->, thick},
  box/.style={
    draw,
    rounded corners,
    inner sep=6pt,
    align=left,
    font=\footnotesize
  }
]

\forestset{
  bubbletree/.style={
    for tree={
      l sep=6pt,
      s sep=2pt,      
      font=\small
    }
  }
}

\node[anchor=north west] (T1) at (0,0) {
  \begin{forest}
    bubbletree
    [\textit{stmt}
      [\term{if}]
      [\term{a}]
      [{\term{=}}][{\term{=}}]
      [\term{b}]
      [\term{then}]
      [\term{x}][{\term{=}}][\term{true}]
      [\term{else}]
      [\term{x}][{\term{=}}][\term{1}][\term{+}][\term{1}][\term{end}]
    ]
  \end{forest}
};

\node (B1) [box, right=6mm of T1] {
  \textbf{Candidates}\\[1mm]
  \textbf{1: \term{1+1}}\\
  2: \term{a==b}\\
  3: \term{x=true}\\
  4: \term{b then x}\\
  \dots
};

\node (T2) [right=6mm of B1] {
  \begin{forest}
    bubbletree
    [\textit{stmt}
      [\term{if}]
      [\term{a}]
      [{\term{=}}][{\term{=}}]
      [\term{b}]
      [\term{then}]
      [\term{x}][{\term{=}}][\term{true}]
      [\term{else}]
      [\term{x}][{\term{=}}]
      [\textit{exp}, draw, circle
        [\textit{exp}, draw, circle[\term{1}]][\term{+}][\textit{exp}, draw, circle[\term{1}]]
      ][\term{end}]
    ]
  \end{forest}
};

\draw[flow] (T1.east) -- (B1.west);
\draw[flow] (B1.east) -- (T2.west);

\node[anchor=north west] (T4) at ($(T1.south west)+(0,-6mm)$) {
  \begin{forest}
    bubbletree
        [\textit{stmt}, draw, circle
          [\term{if}]
          [\textit{bool}
            [\term{a}][{\term{=}}][{\term{=}}][\term{b}]
          ]
          [\term{then}]
          [\textit{stmt}, draw, circle
            [\term{x}][{\term{=}}][\textit{bool}[\term{true}]]
          ]
          [\term{else}]
          [\term{x}][{\term{=}}]
          [\textit{exp}
            [\textit{exp}[\term{1}]][\term{+}][\textit{exp}[\term{1}]]
          ][\term{end}]
        ]
  \end{forest}
};

\node (B3) [box, right=6mm of T4] {
  \sout{1: \term{1+1}}\\
  \sout{2: \term{a==b}}\\
  \textbf{3: \term{x=}\emph{bool}}\\
  4: \term{b then x}\\
  \dots
};

\node (T3) [right=6mm of B3] {
  \begin{forest}
    bubbletree
    [\textit{stmt}
      [\term{if}]
      [\textit{bool}, draw, circle
        [\term{a}][{\term{=}}][{\term{=}}][\term{b}]
      ]
      [\term{then}]
      [\term{x}][{\term{=}}][\textit{bool}, draw, circle[\term{true}]]
      [\term{else}]
      [\term{x}][{\term{=}}]
      [\textit{exp}
        [\textit{exp}[\term{1}]][\term{+}][\textit{exp}[\term{1}]]
      ][\term{end}]
    ]
  \end{forest}
};

\node (B2) [box, right=6mm of T3] {
  \sout{1: \term{1+1}}\\
  \textbf{2: \term{a==b}}\\
  3: \term{x=true}\\
  4: \term{b then x}\\
  \dots
};

\draw[flow] (T2.east) -- ++(2mm,0) -| ([xshift=8mm]B2.north);

\draw[flow] (B2.west) -- (T3.east);
\draw[flow] (T3.west) -- (B3.east);
\draw[flow] (B3.west) -- (T4.east);

\end{tikzpicture}

\caption{Example dynamic batch processing for Lua (whitespace omitted for brevity): 
Initial flat parse tree (top left) with ranked candidate bubbles; 
merging \term{1+1} with \term{1} into \textit{exp};
merging \term{a==b} (kept as spatially disjoint from \term{1+1} merge) with \term{true} into \textit{bool}; 
merging \term{x=}\textit{bool} (after relabeling consumed \term{true})
with \textit{stmt}; 
discarding \term{b then x} (as invalidated by preceding merges).
}
\label{fig:batch-bubble}

\end{figure*}
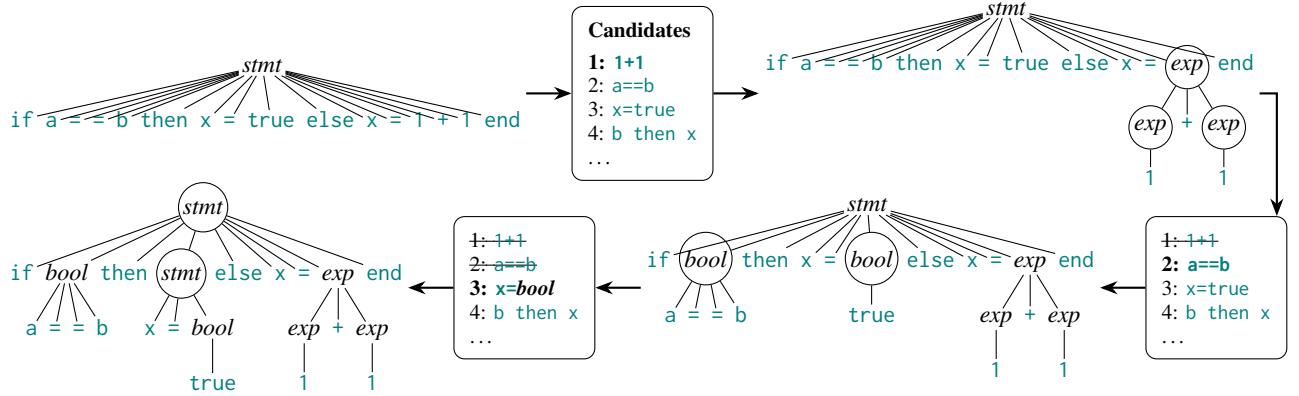

Each successful bubble merge in \arvada{} or \treevada{} has a high overhead, as \arvada{}/\treevada{} update their parse trees, which invalidates indices and the context of the remaining bubble candidates. The approaches thus discard their entire bubble ranking, re-identify all possible bubbles, and re-rank them from scratch. In the Figure~\ref{fig:batch-bubble} example \arvada{}/\treevada{} would discard their ranked candidate bubble list after merging \term{1+1} with \term{1} into \textit{exp}.

We observe that a bubble merge may leave some tree regions unaffected. Instead of discarding the remaining bubble candidates, \toolName{}'s dynamic batch processing keeps attempting to merge the list's bubbles (in rank-order), by evaluating each candidate against the current tree state as follows.

\paragraph{Unaffected (Keep)}
The candidate bubble targets a sequence of nodes that is spatially disjoint from the regions modified by previous merges in the current batch, such as \term{a==b} in Figure~\ref{fig:batch-bubble}. We retain such a candidate as is and attempt to merge it, as it remains strictly valid.

\paragraph{Relabeled (Update)} 
A prior merge in this batch has renamed a node in this candidate bubble. In the Figure~\ref{fig:batch-bubble} example, the \term{a==b} candidate bubble's new parent \textit{bool} merged with the (implicit) parent of \term{true}, which affects the next candidate bubble \term{x=true}. \toolName{} propagates the new label to such a candidate bubble, updating the bubble's elements to match the current tree state while preserving the intended structural grouping. In Figure~\ref{fig:batch-bubble} we relabel \term{x=true} to \term{x=}\textit{bool}.

\paragraph{Broken (Discard)} 
A prior merge has consumed, split, or structurally reconfigured the nodes targeted by this candidate bubble. Due to this structural overlap or conflict, we deem this candidate ``broken'' and discard it. In Figure~\ref{fig:batch-bubble} two merges overlap with the \term{b then x} candidate bubble, which breaks the bubble, as the bubble's nodes are no longer tree siblings.

\subsection{Hierarchical Delta Debugging (HDD) to Decompose Trees} %

\arvada{}/\treevada{} generalize one parse tree per seed program, which can be remarkably effective for small diverse seed programs. But with increasing program size each seed program tends to become a larger monolith with redundancies within and across programs, where each program uses a similar complex structure and no seed program demonstrates simple grammar rule alternatives. This becomes a problem for \arvada{}/\treevada{} if such rules are non-recursive. For example, if each Java seed program declares its superclass then \arvada{}/\treevada{} do not learn the base case of no declared superclass.

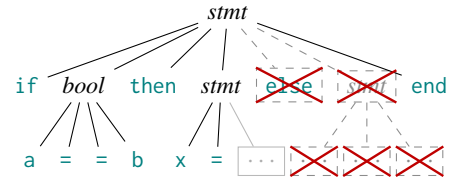
\begin{figure}[h!t]
    \centering
    \begin{forest}
for tree={
  l sep=6pt,
  s sep=2pt,  
  font=\small,
  edge={-},
},
ctx/.style={
  draw=gray!55,
  text=gray!65,
  edge={gray!55},
},
pruned/.style={
  draw=gray!90,
  text=gray!90,
  dashed,
  edge={gray!90,dashed},
},
tikz+={
  \foreach \n in {nelse,nx,neq,nexp,nstmt}{
    \draw[red!75!black, line width=.9pt]
      (\n.north west) -- (\n.south east);
    \draw[red!75!black, line width=.9pt]
      (\n.south west) -- (\n.north east);
  }
}
[\textit{stmt}
  [\term{if}]
  [\textit{bool}
    [\term{a}][{\term{=}}][{\term{=}}][\term{b}]
  ]
  [\term{then}]
  [\textit{stmt}
    [\term{x}][{\term{=}}][{\(\cdots\)}, ctx]
  ]
  [\term{else}, name=nelse, pruned]
  [\textit{stmt}, name=nstmt, pruned
      [{\(\cdots\)}, name=nx,    pruned]
      [{\(\cdots\)},  name=neq,   pruned]
      [{\(\cdots\)}, name=nexp, pruned]
  ]
  [\term{end}]
]
\end{forest}
    \caption{HDD decomposition to reduce a Lua parse tree. 
    }
    \label{fig:hdd}
\end{figure}

In the Figure~\ref{fig:hdd} example, HDD prunes the \texttt{else} branch from the \texttt{if-then-else} construct. After decomposition, the pruned tree induces an additional grammar rule that admits \texttt{if-then} constructs without an \texttt{else} branch, thus exposing a smaller reusable construct that was not present in the seed programs.

Following the idea of hierarchical delta-debugging (HDD)~\cite{hdd,hddzeller}, after bubble exploration we traverse each parse tree top-down from the root. At each visited node $n$, we perform delta-debugging (\textsc{ddmin}~\cite{hddzeller}) on $n$'s child nodes and partition them into $g$ contiguous chunks (starting with $g$ = 2). For each chunk, we construct two candidate trees. One deletes the chunk (and all transitive children). The other deletes $n$'s other (non-chunk) children (and their transitive children). If no candidate tree passes the below oracle check at the current granularity, we increase $g$ (up to the number of $n$'s child nodes) and repeat.

Similar to a bubble merge, we add a candidate tree to our trees $\mathcal{T}$, if our black-box parser accepts 50~sampled programs as valid. We first select the up to 20 trees $\mathcal{T}_t \subseteq \mathcal{T}$ that share the most non-terminals with candidate tree $t$, i.e., we count the elements in the intersection $|\mathit{NT}(t) \cap \mathit{NT}(u)|$ of $t$'s non-terminals $\mathit{NT}(t)$ with another tree $u \in \mathcal{T}$. We sample 50~programs from $\{t\}\cup\mathcal{T}_t$ (up to a rule-expansion depth 2).

\subsection{Descriptive Label for Each Non-terminal via LLM}

Once no further generalization is possible, \toolName{}'s iterative bubble-and-merge process terminates. We then allow \toolName{} to optionally relabel the tree non-terminals so that the tree-induced grammar becomes human comprehensible and thus easier to inspect. For each non-terminal we query a lightweight LLM (GPT 4o-mini) to summarize up to two strings derivable from that non-terminal, as shown in Figure~\ref{fig:prompt_label}.

\begin{figure}[h!t]
\centering
\begin{tcolorbox}[
  width=\columnwidth,
  colback=gray!5!white,
  colframe=gray!75!black,
  boxsep=1mm,
  left=1mm,
  right=1mm,
  top=1mm,
  bottom=1mm
]

\begin{Verbatim}[
  fontsize=\scriptsize,
  breaklines=true,
  breakanywhere=true
]
You are an AI assistant. You will label the internal nodes in the 
parse tree of any arbitrary program. The root node has label stmt. 
You will be given a pair of substrings derivable from the same 
non-terminal symbol. Your job is to label the non-terminal 
...

# Examples:
    - '(n+n)', 'n'
    -> label: numexpr
    ...
\end{Verbatim}
\end{tcolorbox}
\caption{LLM prompt template for descriptive labels.}
\label{fig:prompt_label}
\end{figure}

\subsection{Generalizing Terminals from Seed Observations}

While HDD decomposition yields the final parse trees, the trees' leaf nodes still only reflect the terminal samples we observed in the seed programs (e.g., \term{5} and \term{9} for \textit{num} in Figure~\ref{fig:half-merge}). To approximate the language's likely greater terminal variety, \toolName{} generalizes terminals beyond the observed samples. We thus incrementally assign each terminal (such as \textit{num}) a larger character class, sample 10 programs using the new class, and accept the generalization if the black-box parser accepts all sampled programs.

For alphabetic and numeric tokens \toolName{} follows the  \arvada{}/\treevada{} character-class hierarchy. Both upper-case and lower-case sequences generalize to a mixed-case sequence, which generalizes to an alphanumeric (lower/upper case and positive int) sequence. Similarly, digit $\to$ positive int $\to$ alphanumeric. While \arvada{}/\treevada{} stop there, we then iteratively try to add special characters
\term{!}%
\term{"}%
\term{\#}%
\term{\$}%
\term{\%}%
\term{\&}%
\term{'}%
\term{(}%
\term{)}%
\term{*}%
\term{+}%
\term{,}%
\term{-}%
\term{.}%
\term{/}%
\term{:}%
\term{;}%
\term{<}%
\term{=}%
\term{>}%
\term{?}%
\term{@}%
\term{[}%
\term{\textbackslash}%
\term{]}%
\term{\textsuperscript{$\wedge$}}%
\term{\_}%
\term{`}%
\term{\{}%
\term{|}%
\term{\}}%
\term{\textasciitilde{}}
(one at a time) to the inferred class. Per character, we sample 10 programs and again use the black-box parser as the oracle.

To a quote-bracketed string we apply the same special character generalization, followed by a similar approach to adding whitespace
\term{\textvisiblespace{}}%
\term{\textbackslash{}t}%
\term{\textbackslash{}r}%
\term{\textbackslash{}n}
(one at a time), which we again validate with 10~sampled programs each.

Since \toolName{} infers trees from whitespace-minimized token sequences, the final grammar includes the minimum number of whitespace required for syntactic validity. To support both whitespace-sensitive and -insensitive languages the evaluation parser compiled from the inferred grammar ignores additional whitespace. Note that we do not post process the inferred grammar to remove ambiguities (i.e. fix left recursions). We therefore, use Earley parsing for compiling the evaluation parser, as it can handle any context-free grammar. \toolName{} also allows users to export grammars to the more performant ANTLR4 format, which requires an additional normalization phase to remove indirect left recursions.

\lstdefinelanguage{Grammar}{
  morekeywords={start,stmt,numexpr, boolexpr, n1, n2, n0, n3, n4, t0, t1, stmt_1, stmt_2, boolexpr_, expr, expr_1, expr_2, expr_stmt, t9869, t9562, t7115, t4197, t9770, t4806, t9228, t8994, t6716, t5816, t6146, t5491, t3582, t3830, t1408, t363, t612, t489, t3953, t3325, t2781, t1144, t9342, simple_stmt, assignment, compound_stmt, if_stmt, while_stmt, condition, stmt_13, complex_stmt, conditional_stmt, stmt_4, bool_expr_1, boolstmt, stmt_11, stmt_7, stmt_3, boolstmt_2, complex_numexpr_2, complex_numexpr, complex_expr, unary_expr, unary_op, stmt_15 },    %
  keywordstyle= \normalcolor\itshape, %
  sensitive=true
}

\section{Evaluation Setup}
\label{sec:eval}

We address the following research questions:

\begin{itemize}[noitemsep]
\item \RQAccuracy{}: \RQAccuracyFull{} 
\item \RQCompact: \RQCompactFull{}
\item \RQAblation{}: \RQAblationFull{} 
\item \RQFuzzing{}: \RQFuzzingFull{}
\end{itemize}

\subsection{Languages, Grammars, Parsers, Seed \& Test Programs}
For each language in our study, we obtain its ``golden'' grammar (where available), valid seed programs (Table~\ref{table:seed_stat}), an existing ``golden'' black-box parser and ``golden'' valid test programs from the following grammar-mining artifacts and grammar repositories. We use \textit{turtle}, \textit{while}, \textit{lisp}, \textit{xml}, \textit{json}, \textit{tinyc}, \textit{c-500}, and \textit{curl} from \treevada{}~\cite{Arefin2023}. (\treevada{} subjects use R1 seed variants, \emph{c-500} is \emph{tinyc} with the larger R5 seeds.) We add \textit{minic} from \kedavra{}\footnote{\Accessed \url{https://github.com/Sinpersrect/kedavra}} and \textit{tiny}, \textit{lua}, \textit{c}, and \textit{mysql} from \gladeRep{}~\cite{bachir_bendrissou_artifact}. We further add \textit{liquid}, using Python Liquid\footnote{\Accessed \url{https://github.com/jg-rp/liquid/tree/v2.2.1}} as the black-box parser and collecting its seed programs and 1k~held-out test programs from Shopify's \textit{liquid-spec} suite\footnote{\Accessed \url{https://github.com/Shopify/liquid-spec}}. Finally, we add \textit{java}, \textit{cpp}, and \textit{rust} grammars from the ANTLR4 grammar repository\footnote{\Accessed \url{https://github.com/antlr/grammars-v4}} and generate their seed/test programs the same way as for the \gladeRep{} subjects below.

\nprounddigits{1}
\begin{table*}[h!t]
\centering
\caption{Lower- (top) and higher-complexity (bottom) golden grammars/languages (left) and their seed programs (right);
T~=~terminals; 
NT~=~non-terminals; 
RHS~=~avg. right-hand side length; 
MCC~=~avg. McCabe cyclomatic complexity;
D~=~largest minimum NT distance from \emph{start};
LRS~=~LR automaton states;
LAT~=~lookahead metric; 
LTPSM~=~max terminal pairs;
P~=~programs;
C\textsubscript{avg}/C\textsubscript{max}~=~avg/max characters; 
T\textsubscript{avg}/T\textsubscript{max}~=~avg/max tokens; 
B~=~pre-structured trees' branch factor.
}
\label{table:seed_stat}
\begin{tabular}{l|rrN{2}{1}N{2}{1}r|rN{3}{1}r|r|N{3}{1}r|N{3}{1}r|N{2}{1}}
\toprule
 & \multicolumn{5}{c}{Golden Grammar} & \multicolumn{3}{c|}{Parser \& Language} & \multicolumn{6}{c}{Seed Programs}  \\
 & \multicolumn{1}{c}{T} 
 & \multicolumn{1}{c}{NT} 
 & \multicolumn{1}{c}{RHS} 
 & \multicolumn{1}{c}{MCC}
 & \multicolumn{1}{c|}{D}
 & \multicolumn{1}{c}{LRS} 
 & \multicolumn{1}{c}{LAT} 
 & \multicolumn{1}{c|}{LTPSM} 
 & \multicolumn{1}{c|}{P} 
 & \multicolumn{1}{c}{C\textsubscript{avg}} 
 & \multicolumn{1}{c|}{C\textsubscript{max}} 
 & \multicolumn{1}{c}{T\textsubscript{avg}} 
 & \multicolumn{1}{c|}{T\textsubscript{max}} 
 & \multicolumn{1}{c}{B} \\
\midrule
 \Use{turtle_programname}  & \Use{turtle_grammar_stat_term} & \Use{turtle_grammar_stat_non_term} & \Use{turtle_grammar_rhs_length} & \Use{turtle_grammar_mcc} & \Use{turtle_grammar_ntd} & \Use{turtle_grammar_lrs} & \Use{turtle_grammar_lat} & \Use{turtle_grammar_ltpsm}  & \Use{turtle_program_seed_no}  & \Use{turtle_avg_char} & \Use{turtle_max_char} & \Use{turtle_avg_tokens} & \Use{turtle_max_tokens} & \Use{turtle_branching_factor}\\
 \Use{while_programname}  & \Use{while_grammar_stat_term} & \Use{while_grammar_stat_non_term} & \Use{while_grammar_rhs_length} & \Use{while_grammar_mcc} & \Use{while_grammar_ntd} & \Use{while_grammar_lrs} & \Use{while_grammar_lat} & \Use{while_grammar_ltpsm} & \Use{while_program_seed_no}  & \Use{while_avg_char} & \Use{while_max_char} & \Use{while_avg_tokens} & \Use{while_max_tokens} & \Use{while_branching_factor}\\
 \Use{lisp_programname}  & \Use{lisp_grammar_stat_term} & \Use{lisp_grammar_stat_non_term} & \Use{lisp_grammar_rhs_length} & \Use{lisp_grammar_mcc} & \Use{lisp_grammar_ntd} & \Use{lisp_grammar_lrs} & \Use{lisp_grammar_lat} & \Use{lisp_grammar_ltpsm} & \Use{lisp_program_seed_no}  & \Use{lisp_avg_char} & \Use{lisp_max_char} & \Use{lisp_avg_tokens} & \Use{lisp_max_tokens} & \Use{lisp_branching_factor}\\
 \Use{xml_programname}  & \Use{xml_grammar_stat_term} & \Use{xml_grammar_stat_non_term} & \Use{xml_grammar_rhs_length} & \Use{xml_grammar_mcc} & \Use{xml_grammar_ntd} & \Use{xml_grammar_lrs} & \Use{xml_grammar_lat} & \Use{xml_grammar_ltpsm} & \Use{xml_program_seed_no}  & \Use{xml_avg_char} & \Use{xml_max_char} & \Use{xml_avg_tokens} & \Use{xml_max_tokens} & \Use{xml_branching_factor}\\
 \Use{json_programname}  & \Use{json_grammar_stat_term} & \Use{json_grammar_stat_non_term} & \Use{json_grammar_rhs_length} & \Use{json_grammar_mcc} & \Use{json_grammar_ntd} & \Use{json_grammar_lrs} & \Use{json_grammar_lat} & \Use{json_grammar_ltpsm} & \Use{json_program_seed_no}  & \Use{json_avg_char} & \Use{json_max_char} & \Use{json_avg_tokens} & \Use{json_max_tokens} & \Use{json_branching_factor}\\
 \Use{tinyc_programname}  & \Use{tinyc_grammar_stat_term} & \Use{tinyc_grammar_stat_non_term} & \Use{tinyc_grammar_rhs_length} & \Use{tinyc_grammar_mcc} & \Use{tinyc_grammar_ntd} & \Use{tinyc_grammar_lrs} & \Use{tinyc_grammar_lat} & \Use{tinyc_grammar_ltpsm} & \Use{tinyc_program_seed_no}  & \Use{tinyc_avg_char} & \Use{tinyc_max_char} & \Use{tinyc_avg_tokens} & \Use{tinyc_max_tokens} & \Use{tinyc_branching_factor}\\
 \Use{c-500_programname}  & \Use{c-500_grammar_stat_term} & \Use{c-500_grammar_stat_non_term} & \Use{c-500_grammar_rhs_length} & \Use{c-500_grammar_mcc} & \Use{tinyc_grammar_ntd} & \Use{c-500_grammar_lrs} & \Use{c-500_grammar_lat} & \Use{c-500_grammar_ltpsm} & \Use{c-500_program_seed_no}  & \Use{c-500_avg_char} & \Use{c-500_max_char} & \Use{c-500_avg_tokens} & \Use{c-500_max_tokens} & \Use{c-500_branching_factor}\\
 \Use{tiny_programname}  & \Use{tiny_grammar_stat_term} & \Use{tiny_grammar_stat_non_term} & \Use{tiny_grammar_rhs_length} & \Use{tiny_grammar_mcc} & \Use{tiny_grammar_ntd} & \Use{tiny_grammar_lrs} & \Use{tiny_grammar_lat} & \Use{tiny_grammar_ltpsm} & \Use{tiny_program_seed_no}  & \Use{tiny_avg_char} & \Use{tiny_max_char} & \Use{tiny_avg_tokens} & \Use{tiny_max_tokens} & \Use{tiny_branching_factor}\\
 \Use{minic_programname}  & \Use{minic_grammar_stat_term} & \Use{minic_grammar_stat_non_term} & \Use{minic_grammar_rhs_length} & \Use{minic_grammar_mcc} & \Use{minic_grammar_ntd} & \Use{minic_grammar_lrs} & \Use{minic_grammar_lat} & \Use{minic_grammar_ltpsm} & \Use{minic_program_seed_no}  & \Use{minic_avg_char} & \Use{minic_max_char} & \Use{minic_avg_tokens} & \Use{minic_max_tokens} & \Use{minic_branching_factor}\\
 \Use{curl_programname}  & \Use{curl_grammar_stat_term} & \Use{curl_grammar_stat_non_term} & \Use{curl_grammar_rhs_length} & \Use{curl_grammar_mcc} & \Use{curl_grammar_ntd} & \Use{curl_grammar_lrs} & \Use{curl_grammar_lat} & \Use{curl_grammar_ltpsm} & \Use{curl_program_seed_no}  & \Use{curl_avg_char} & \Use{curl_max_char} & \Use{curl_avg_tokens} & \Use{curl_max_tokens} & \Use{curl_branching_factor}\\
\midrule
 \Use{liquid_programname}  & \Use{liquid_grammar_stat_term} & \Use{liquid_grammar_stat_non_term} & \Use{liquid_grammar_rhs_length} & \Use{liquid_grammar_mcc} & \Use{liquid_grammar_ntd} & \Use{liquid_grammar_lrs} & \Use{liquid_grammar_lat} & \Use{liquid_grammar_ltpsm} & \Use{liquid_program_seed_no}  & \Use{liquid_avg_char} & \Use{liquid_max_char} & \Use{liquid_avg_tokens} & \Use{liquid_max_tokens} & \Use{liquid_branching_factor}\\
\Use{lua_programname}  & \Use{lua_grammar_stat_term} & \Use{lua_grammar_stat_non_term} & \Use{lua_grammar_rhs_length} & \Use{lua_grammar_mcc} & \Use{lua_grammar_ntd} & \Use{lua_grammar_lrs} & \Use{lua_grammar_lat} & \Use{lua_grammar_ltpsm} & \Use{lua_program_seed_no}  & \Use{lua_avg_char} & \Use{lua_max_char} & \Use{lua_avg_tokens} & \Use{lua_max_tokens} & \Use{lua_branching_factor}\\
 \Use{c_programname}  & \Use{c_grammar_stat_term} & \Use{c_grammar_stat_non_term} & \Use{c_grammar_rhs_length} & \Use{c_grammar_mcc} & \Use{c_grammar_ntd} & \Use{c_grammar_lrs} & \Use{c_grammar_lat} & \Use{c_grammar_ltpsm} & \Use{c_program_seed_no}  & \Use{c_avg_char} & \Use{c_max_char} & \Use{c_avg_tokens} & \Use{c_max_tokens} & \Use{c_branching_factor}\\
\Use{java_programname}  & \Use{java_grammar_stat_term} & \Use{java_grammar_stat_non_term} & \Use{java_grammar_rhs_length} & \Use{java_grammar_mcc} & \Use{java_grammar_ntd} & \Use{java_grammar_lrs} & \Use{java_grammar_lat} & \Use{java_grammar_ltpsm} & \Use{java_program_seed_no}  & \Use{java_avg_char} & \Use{java_max_char} & \Use{java_avg_tokens} & \Use{java_max_tokens} & \Use{java_branching_factor}\\
\Use{cpp_programname}  & \Use{cpp_grammar_stat_term} & \Use{cpp_grammar_stat_non_term} & \Use{cpp_grammar_rhs_length} & \Use{cpp_grammar_mcc} & \Use{cpp_grammar_ntd} & \Use{cpp_grammar_lrs} & \Use{cpp_grammar_lat} & \Use{cpp_grammar_ltpsm} & \Use{cpp_program_seed_no}  & \Use{cpp_avg_char} & \Use{cpp_max_char} & \Use{cpp_avg_tokens} & \Use{cpp_max_tokens} & \Use{cpp_branching_factor}\\
\Use{rust_programname}  & \Use{rust_grammar_stat_term} & \Use{rust_grammar_stat_non_term} & \Use{rust_grammar_rhs_length} & \Use{rust_grammar_mcc} & \Use{rust_grammar_ntd} & \Use{rust_grammar_lrs} & \Use{rust_grammar_lat} & \Use{rust_grammar_ltpsm} & \Use{rust_program_seed_no}  & \Use{rust_avg_char} & \Use{rust_max_char} & \Use{rust_avg_tokens} & \Use{rust_max_tokens} & \Use{rust_branching_factor}\\
 \Use{mysql_programname}  & \Use{mysql_grammar_stat_term} & \Use{mysql_grammar_stat_non_term} & \Use{mysql_grammar_rhs_length} & \Use{mysql_grammar_mcc} & \Use{mysql_grammar_ntd} & \Use{mysql_grammar_lrs} & \Use{mysql_grammar_lat} & \Use{mysql_grammar_ltpsm} & \Use{mysql_program_seed_no}  & \Use{mysql_avg_char} & \Use{mysql_max_char} & \Use{mysql_avg_tokens} & \Use{mysql_max_tokens} & \Use{mysql_branching_factor}\\
\bottomrule
\end{tabular}
\end{table*}

\MyPara{Excessive duplicates in \gladeRep{}}
The \gladeRep{} replication study by Bendrissou et al., published at PLDI~2022~\cite{bendrissou2022synthesizing}, made several contributions, including re-implementing the \glade{} tool and releasing new evaluation subjects~\cite{bachir_bendrissou_artifact}. Each \gladeRep{} subject comes with 1k~test programs. But of its 18 ANTLR4 subjects, only \textit{pascal} has no duplicate test program. Most affected is \textit{ints}, which has 139 unique test programs but (among others) 515 test programs that only contain the single character \term{0}. As a consequence, even a trivial grammar that only recognizes \term{0} would score 51.5\% recall on \textit{ints}. Similarly, lua has 462 unique test programs (but 322 programs that only contain a single whitespace character) and in mysql 305 are unique (while 287 are empty).

Such artifacts are not representative of the corresponding golden grammar and thus yield misleading recall and F1 scores. To avoid this issue, we use we use the golden grammars for tiny, lua, c, and mysql from \gladeRep{} and generate new seed and test sets without duplicates using the off-the-shelf grammar fuzzer \emph{Grammarinator}~\cite{hodovan2018grammarinator}. To avoid excessive program growth, we set the fuzzer’s \emph{max-depth} parameter to the minimum depth required to reach every non-terminal from the \emph{start} rule (Table~\ref{table:seed_stat}, column~D).

\subsection{Golden Grammar and Language Complexity: Lo vs. Hi}

Table~\ref{table:seed_stat} orders our golden grammars by non-terminals (NT), which correlates with several other Table~\ref{table:seed_stat} size and complexity metrics\footnote{Metrics computed via gMetrics~\cite{Crepinsek2010} (which extends SynC~\cite{Power2004,Csuhaj-Varju1993}). gMetrics uses ANTLR v3, so we slightly modified v4 grammars to v3 syntax.}. At a high level, higher metric values imply larger grammar size and complexity. We create two groups: small/toy (top) and larger / more complex (bottom).

Basic metrics are number of terminals~(T) and non-terminals~(NT). The average right-hand side length~(RHS) divides total terminal and non-terminal occurrences on all rules' right-hand sides by the number of non-terminals~(NT). The average McCabe cyclomatic complexity~(MCC) divides the total number of ``branch'' operators (alternative rule~\term{|}, optional term~\term{?}, and closures~\term{*} and~\term{+}) of all rules by the number of non-terminals (NT). Depth~(D) is the largest minimum distance of a non-terminal from the grammar's \emph{start} non-terminal.

Two metrics generate from the grammar a LR parser~\cite{Aho2006} and count its properties, e.g., the LR automaton's states (LRS). The basic lookahead metric (LAT) adds for each terminal the number of LR automaton states that do not lead to an error when that terminal is in the lookahead, divided by the number of terminals (T). The final metric captures a language property directly from the grammar rules. The maximum number of different terminal pairs starting with a given terminal (LTPSM) iteratively expands all grammar rules to compute all possible terminal pairs.

By comparing 16~languages (including DSLs and several Java versions) on 21~metrics, gMetrics proposes four grammar categories~\cite[Table~6]{Crepinsek2010}: tiny, small (mainly DSLs), intermediate (3rd-generation general-purpose), and big. The cut-offs are 
LRS (100, 1k, 3k), 
LAT (20, 200, 500), 
and LTPSM (6, 50, 70).
The cut-offs align with our two groups of low (tiny and small) and high (intermediate and big) complexity grammars for our Table~\ref{table:seed_stat} subjects. There are three exceptions. (1)~First, with no golden grammar available, we categorize curl as Lo as it is a regular language. (2)~Second, also with no golden grammar available, we categorize liquid as Hi because its collected programs contain context-free structure, e.g., nested brackets, and the inferred grammar statistics align more closely with the Hi group. (3)~Third, with only partial gMetrics results, we categorize mysql as Hi as it scores highest on the available metrics, i.e., terminals, non-terminals, and depth.

\subsection{Metrics: Runtime, Inferred Grammars' Size \& Accuracy}

The Crucio artifact\footnote{\Accessed \url{https://github.com/Sinpersrect/crucio}} does not generate any grammar file output. To compare with Crucio, we convert its program generation model $P_{\text{sample}}$ into an equivalent set of production rules and export the resulting grammar. To evaluate how accurately an inferred grammar $\mathcal{G}$ captures the target language, we then use the three \arvada{}/\treevada{} study metrics:

\MyPara{Precision} We sample 1k programs from $\mathcal{G}$ and count how many are accepted by the existing (``golden'') black-box parser. 

\MyPara{Recall} We use the existing held-out (``golden'') test set of valid programs (e.g., generated from the golden grammar) and count how many are accepted by a Python Earley parser generated from $\mathcal{G}$.

\MyPara{F1 Score} As usual, F1 is the harmonic mean of precision $p$ and recall $r$: $2*p*r/(p+r)$, which is widely considered a useful overall metric for the inferred grammar's accuracy.

For precision \arvada{} and \treevada{} used a fixed grammar expansion depth of 5 (which does not reach deeper nested non-terminals), while \kedavra{} limits each production rule to at most 10 uses (which may also prevent deeply nested rules from being used). To ensure that our sampling can exercise all reachable non-terminals, our evaluation just sets each grammar's maximum rule expansion depth to the minimum depth required to reach any non-terminal from the \emph{start} rule (i.e., the grammar's Table~\ref{table:stat_table} D value).

\MyPara{Grammar Size} Given two equally accurate grammars, a human consumer may prefer the more compact one. We thus count across all rule alternatives $A$ all terminal and non-terminal occurrences: $S=A*l(A)$, which for the Table~\ref{table:seed_stat} golden grammars is \textit{NT}$*$\textit{RHS}.

\MyPara{Runtime} For Table~\ref{table:main_table} we ran each experiment with a 48h time-out on a 2.1GHz Intel Xeon Platinum 8160 machine with 192~GB RAM running Rocky Linux 9.7. To approximate a user's experience with the non-deterministic subjects, for \kedavra{} and \crucio{} we ran each experiment five times and report the average. Reported runtimes do not include converting \crucio's internal representation to a grammar output file or measuring precision or recall.

\nprounddigits{1}

\section{Results}

Table~\ref{table:main_table} shows grammar accuracy and inference time. Each parse attempt yielded a valid vs. invalid result before our 5~minute time-out. Table~\ref{table:stat_table} characterizes the resulting grammar sizes. We discuss the Lo and Hi language groups separately because the tools exhibit substantially different behavior across these groups.

\begin{table*}[h!t]
\centering
\caption{Runtime \& inferred grammars' precision \& recall (5-run avg. for non-deterministic \kedavra{} and \crucio{}); for the larger lanuguages \kedavra{} and \crucio{} failed, timed out (48h), or yielded no recall;
bold~=~best;
underline~=~runner-up.
}
\label{table:main_table}
\begin{tabular}{l|r r r r| r r r r|r r r r|r r r r}
\toprule
 &  \multicolumn{4}{c|}{\treevada{}} & \multicolumn{4}{c|}{\kedavra{}} &  \multicolumn{4}{c|}{Crucio}  & \multicolumn{4}{c}{\toolName{}} \\
\multicolumn{1}{c|}{}  & \multicolumn{1}{c}{p} & \multicolumn{1}{c}{r} & \multicolumn{1}{c}{f1} & \multicolumn{1}{c|}{t[s]} & \multicolumn{1}{c}{p} & \multicolumn{1}{c}{r} & \multicolumn{1}{c}{f1} & \multicolumn{1}{c|}{t[s]} & \multicolumn{1}{c}{p} & \multicolumn{1}{c}{r} & \multicolumn{1}{c}{f1} & \multicolumn{1}{c|}{t[s]} & \multicolumn{1}{c}{p} & \multicolumn{1}{c}{r} & \multicolumn{1}{c}{f1} & \multicolumn{1}{c}{t[s]} \\
\midrule
 \Use{turtle_programname} & \Use{turtle_treevada_precision}  & \Use{turtle_treevada_recall} & \Use{turtle_treevada_f1} & \Use{turtle_treevada_build_time} & \Use{turtle_Kedavra_precision}  & \Use{turtle_Kedavra_recall} & \textbf{\Use{turtle_Kedavra_f1}} & \Use{turtle_Kedavra_build_time} & \Use{turtle_crucio_precision} & \Use{turtle_crucio_recall} & \Use{turtle_crucio_f1} & \Use{turtle_crucio_build_time} & \Use{turtle_xvada_precision} & \Use{turtle_xvada_recall} & \textbf{\Use{turtle_xvada_f1}} & \Use{turtle_xvada_build_time} \\
 \Use{while_programname} & \Use{while_treevada_precision}  & \Use{while_treevada_recall} & \textbf{\Use{while_treevada_f1}} & \Use{while_treevada_build_time} & \Use{while_Kedavra_precision}  & \Use{while_Kedavra_recall} & \textbf{\Use{while_Kedavra_f1}} & \textbf{\Use{while_Kedavra_build_time}} & \Use{while_crucio_precision} & \Use{while_crucio_recall} & \Use{while_crucio_f1} & \Use{while_crucio_build_time} & \Use{while_xvada_precision} & \Use{while_xvada_recall} & \textbf{\Use{while_xvada_f1}} & \Use{while_xvada_build_time} \\
 \Use{lisp_programname} & \Use{lisp_treevada_precision}  & \Use{lisp_treevada_recall} & \Use{lisp_treevada_f1} & \Use{lisp_treevada_build_time} & \Use{lisp_Kedavra_precision}  & \Use{lisp_Kedavra_recall} & \textbf{\Use{lisp_Kedavra_f1}} & \textbf{\Use{lisp_Kedavra_build_time}} & \Use{lisp_crucio_precision} & \Use{lisp_crucio_recall} & \Use{lisp_crucio_f1} & \Use{lisp_crucio_build_time} & \Use{lisp_xvada_precision} & \Use{lisp_xvada_recall} & \textbf{\Use{lisp_xvada_f1}} & \Use{lisp_xvada_build_time} \\
 \Use{xml_programname} & \Use{xml_treevada_precision}  & \Use{xml_treevada_recall} & \textbf{\Use{xml_treevada_f1}} & \textbf{\Use{xml_treevada_build_time}} & \Use{xml_Kedavra_precision}  & \Use{xml_Kedavra_recall} & \textbf{\Use{xml_Kedavra_f1}} & \Use{xml_Kedavra_build_time} & \Use{xml_crucio_precision} & \Use{xml_crucio_recall} & \Use{xml_crucio_f1} & \Use{xml_crucio_build_time} & \Use{xml_xvada_precision} & \Use{xml_xvada_recall} & \Use{xml_xvada_f1} & \Use{xml_xvada_build_time} \\
 \Use{json_programname} & \Use{json_treevada_precision}  & \Use{json_treevada_recall} & \Use{json_treevada_f1} & \Use{json_treevada_build_time} & \Use{json_Kedavra_precision}  & \Use{json_Kedavra_recall} & \textbf{\Use{json_Kedavra_f1}} & \Use{json_Kedavra_build_time} & \Use{json_crucio_precision} & \Use{json_crucio_recall} & \Use{json_crucio_f1} & \Use{json_crucio_build_time} & \Use{json_xvada_precision} & \Use{json_xvada_recall} & \underline{\Use{json_xvada_f1}} & \Use{json_xvada_build_time} \\
 \Use{tinyc_programname} & \Use{tinyc_treevada_precision}  & \Use{tinyc_treevada_recall} & \underline{\Use{tinyc_treevada_f1}} & \Use{tinyc_treevada_build_time} & \Use{tinyc_Kedavra_precision}  & \Use{tinyc_Kedavra_recall} & \textbf{\Use{tinyc_Kedavra_f1}} & \textbf{\Use{tinyc_Kedavra_build_time}} & \Use{tinyc_crucio_precision} & \Use{tinyc_crucio_recall} & \Use{tinyc_crucio_f1} & \Use{tinyc_crucio_build_time} & \Use{tinyc_xvada_precision} & \Use{tinyc_xvada_recall} & \Use{tinyc_xvada_f1} & \Use{tinyc_xvada_build_time} \\
 \Use{c-500_programname} & \Use{c-500_treevada_precision}  & \Use{c-500_treevada_recall} & \Use{c-500_treevada_f1} & \Use{c-500_treevada_build_time} & \Use{c-500_Kedavra_precision}  & \Use{c-500_Kedavra_recall} & \textbf{\Use{c-500_Kedavra_f1}} & \textbf{\Use{c-500_Kedavra_build_time}} & \Use{c-500_crucio_precision} & \Use{c-500_crucio_recall} & \Use{c-500_crucio_f1} & \Use{c-500_crucio_build_time} & \Use{c-500_xvada_precision} & \Use{c-500_xvada_recall} & \underline{\Use{c-500_xvada_f1}} & \Use{c-500_xvada_build_time} \\
 \Use{tiny_programname} & \Use{tiny_treevada_precision}  & \Use{tiny_treevada_recall} & \underline{\Use{tiny_treevada_f1}} & \Use{tiny_treevada_build_time} & \Use{tiny_Kedavra_precision}  & \Use{tiny_Kedavra_recall} & \Use{tiny_Kedavra_f1} & \Use{tiny_Kedavra_build_time} & \Use{tiny_crucio_precision} & \Use{tiny_crucio_recall} & \Use{tiny_crucio_f1} & \Use{tiny_crucio_build_time} & \Use{tiny_xvada_precision} & \Use{tiny_xvada_recall} & \textbf{\Use{tiny_xvada_f1}} & \Use{tiny_xvada_build_time} \\
 \Use{minic_programname} & \Use{minic_treevada_precision}  & \Use{minic_treevada_recall} & \Use{minic_treevada_f1} & \Use{minic_treevada_build_time} & \Use{minic_Kedavra_precision}  & \Use{minic_Kedavra_recall} & \underline{\Use{minic_Kedavra_f1}} & \Use{minic_Kedavra_build_time} & \Use{minic_crucio_precision} & \Use{minic_crucio_recall} & \Use{minic_crucio_f1} & \Use{minic_crucio_build_time} & \Use{minic_xvada_precision} & \Use{minic_xvada_recall} & \textbf{\Use{minic_xvada_f1}} & \Use{minic_xvada_build_time} \\
 \Use{curl_programname} & \Use{curl_treevada_precision}  & \Use{curl_treevada_recall} & \textbf{\Use{curl_treevada_f1}} & \Use{curl_treevada_build_time} & \Use{curl_Kedavra_precision}  & \Use{curl_Kedavra_recall} & \Use{curl_Kedavra_f1} & \Use{curl_Kedavra_build_time} & \Use{curl_crucio_precision} & \Use{curl_crucio_recall} & \Use{curl_crucio_f1} & \Use{curl_crucio_build_time} & \Use{curl_xvada_precision} & \Use{curl_xvada_recall} & \underline{\Use{curl_xvada_f1}} & \Use{curl_xvada_build_time} \\
\midrule
 \Use{Avg_lo_programname} & \Use{Avg_lo_treevada_precision}  & \Use{Avg_lo_treevada_recall} & \underline{\Use{Avg_lo_treevada_f1}} & \Use{Avg_lo_treevada_build_time} & \Use{Avg_lo_Kedavra_precision}  & \Use{Avg_lo_Kedavra_recall} & \Use{Avg_lo_Kedavra_f1} & \textbf{\Use{Avg_lo_Kedavra_build_time}} & \Use{Avg_lo_crucio_precision} & \Use{Avg_lo_crucio_recall} & \Use{Avg_lo_crucio_f1} & \Use{Avg_lo_crucio_build_time} & \Use{Avg_lo_xvada_precision} & \Use{Avg_lo_xvada_recall} & \textbf{\Use{Avg_lo_xvada_f1}} & \Use{Avg_lo_xvada_build_time} \\
\midrule
\Use{liquid_programname} & \Use{liquid_treevada_precision}  & \Use{liquid_treevada_recall} & \underline{\Use{liquid_treevada_f1}} & \Use{liquid_treevada_build_time} & \multicolumn{3}{c}{n/a} & \Use{liquid_Kedavra_build_time} & \Use{liquid_crucio_precision} & \Use{liquid_crucio_recall} & \Use{liquid_crucio_f1} & \Use{liquid_crucio_build_time} & \Use{liquid_xvada_precision} & \Use{liquid_xvada_recall} & \textbf{\Use{liquid_xvada_f1}} & \Use{liquid_xvada_build_time} \\
\Use{lua_programname} & \Use{lua_treevada_precision}  & \Use{lua_treevada_recall} & \underline{\Use{lua_treevada_f1}} & \Use{lua_treevada_build_time} & \multicolumn{3}{c}{n/a} & \Use{lua_Kedavra_build_time} & \Use{lua_crucio_precision} & \Use{lua_crucio_recall} & \Use{lua_crucio_f1} & \Use{lua_crucio_build_time} & \Use{lua_xvada_precision} & \Use{lua_xvada_recall} & \textbf{\Use{lua_xvada_f1}} & \Use{lua_xvada_build_time} \\
 \Use{c_programname} & \Use{c_treevada_precision}  & \Use{c_treevada_recall} & \underline{\Use{c_treevada_f1}} & \Use{c_treevada_build_time} & \multicolumn{3}{c}{n/a} & \Use{c_Kedavra_build_time} & \Use{c_crucio_precision} & \Use{c_crucio_recall} & \Use{c_crucio_f1} & \Use{c_crucio_build_time} & \Use{c_xvada_precision} & \Use{c_xvada_recall} & \textbf{\Use{c_xvada_f1}} & \Use{c_xvada_build_time} \\ 
 \Use{java_programname} & \Use{java_treevada_precision}  & \Use{java_treevada_recall} & \underline{\Use{java_treevada_f1}} & \Use{java_treevada_build_time} & \multicolumn{3}{c}{n/a} & \Use{java_Kedavra_build_time} & \Use{java_crucio_precision} & \Use{java_crucio_recall} & \Use{java_crucio_f1} & \Use{java_crucio_build_time} & \Use{java_xvada_precision} & \Use{java_xvada_recall} & \textbf{\Use{java_xvada_f1}} & \textbf{\Use{java_xvada_build_time}} \\
 \Use{cpp_programname} & \Use{cpp_treevada_precision}  & \Use{cpp_treevada_recall} & \underline{\Use{cpp_treevada_f1}} & \Use{cpp_treevada_build_time} & \multicolumn{3}{c}{n/a} & \Use{cpp_Kedavra_build_time} & \multicolumn{3}{c}{n/a} & \Use{cpp_crucio_build_time} & \Use{cpp_xvada_precision} & \Use{cpp_xvada_recall} & \textbf{\Use{cpp_xvada_f1}} & \Use{cpp_xvada_build_time} \\
 \Use{rust_programname} & \Use{rust_treevada_precision}  & \Use{rust_treevada_recall} & \underline{\Use{rust_treevada_f1}} & \Use{rust_treevada_build_time} & \multicolumn{3}{c}{n/a} & \Use{rust_Kedavra_build_time} & \multicolumn{3}{c}{n/a} & \Use{rust_crucio_build_time} & \Use{rust_xvada_precision} & \Use{rust_xvada_recall} & \textbf{\Use{rust_xvada_f1}} & \textbf{\Use{rust_xvada_build_time}} \\

 \Use{mysql_programname} & \Use{mysql_treevada_precision}  & \Use{mysql_treevada_recall} & \underline{\Use{mysql_treevada_f1}} & \Use{mysql_treevada_build_time} & \multicolumn{3}{c}{n/a} & \Use{mysql_Kedavra_build_time} & \multicolumn{3}{c}{n/a} & \Use{mysql_crucio_build_time} & \Use{mysql_xvada_precision} & \Use{mysql_xvada_recall} & \textbf{\Use{mysql_xvada_f1}} & \Use{mysql_xvada_build_time} \\
 \midrule
 \Use{Avg_hi_programname} & \Use{Avg_hi_treevada_precision}  & \Use{Avg_hi_treevada_recall} & \underline{\Use{Avg_hi_treevada_f1}} & \Use{Avg_hi_treevada_build_time} & \multicolumn{4}{c|}{n/a} & \multicolumn{4}{c|}{n/a} & \Use{Avg_hi_xvada_precision} & \Use{Avg_hi_xvada_recall} & \textbf{\Use{Avg_hi_xvada_f1}} & \Use{Avg_hi_xvada_build_time} \\
\bottomrule
\end{tabular}
\end{table*}

\subsection{\RQAccuracy{}: Effectiveness}  %

\MyPara{\kedavra{} on Lo languages}
\kedavra{} was highly effective on the seven Lo subjects whose golden grammars have fewer than ten non-terminals~(Table~\ref{table:seed_stat}): it obtains an F1 score of 1.00 on six subjects and 0.98 on \textit{json}. It is also the fastest tool on four of these subjects (\textit{while}, \textit{lisp}, \textit{tinyc}, and \textit{c-500}). Its performance is weaker on the remaining Lo subjects with larger golden grammars: F1 falls to 0.67 on \textit{tiny} (13 non-terminals) and 0.65 on \textit{minic} (14 non-terminals), and to 0.25 on \textit{curl}. Overall, \kedavra{} attains an average Lo-group F1 of 0.85 in 198 seconds, compared with \Tool{}'s 0.88 in 485 seconds and \treevada{}'s 0.87 in 1,134 seconds. However, this pattern does not extend to the Hi group: \kedavra{} terminates with an unhandled exception during inference for \textit{liquid}, \textit{c} and \textit{rust}, and does not infer a grammar for \textit{lua}, \textit{java}, \textit{cpp}, or \textit{mysql} within the 48-hour limit.

\MyPara{\crucio{} under a fair evaluation protocol}
\crucio{} completes all Lo subjects and four Hi subjects (\textit{liquid}, \textit{lua}, \textit{c}, and \textit{java}). However, when its exported grammar is evaluated using the same full-program sampling and parsing protocol as the other tools (Section~\ref{sec:eval}), its F1 scores are extremely low: its best result is 0.47 on \textit{while}, while all other completed subjects score at most 0.09. The \textit{while} golden grammar uses only the fixed terminals \term{L} and \term{n} at expression positions rather than a general identifier rule. This result is consistent with the evaluation asymmetry discussed in Section~\ref{sec:methodology}.

\MyPara{\toolName{} across both groups}
\Tool{} achieves the highest average F1 score in the Lo group (88\%), narrowly exceeding \treevada{} (87\%) and \kedavra{} (85\%). While these accuracy scores are comparable, \Tool{} reduces the average grammar-construction time by 57\% relative to \treevada{} (485 versus 1,134 seconds); \kedavra{} is faster still where it completes (198 seconds). In the Hi group, \treevada{} infers a grammar for all seven languages but obtains F1 scores between 4~and 26\%. \Tool{} improves the F1 score on every Hi subject, with its strongest result of 99\% on \textit{liquid}, and raises the group average from 12~to 47\%. \Tool{} thus provides the strongest accuracy among the tools that complete the Hi-group experiments, although its average runtime is 9\% higher than \treevada{}'s (51k vs. 47k seconds).

\begin{framed}
\noindent
\emph{
\textbf{\RQAccuracy{}:}
On Lo, \Tool{} matches the best baseline's (\treevada{}) accuracy (at faster average runtime). On Hi, only \treevada{} and \Tool{} succeed on every experiment; \Tool{} improves \treevada{}'s average F1 score by 35 percentage points, from 12 to 47\%.
}
\end{framed}

\subsection{\RQCompact{}: Grammar Size (Compactness)}

EBNF grammar representation is typically more compact than BNF. As \kedavra{} counts EBNF elements (as opposed to unfolding \term{?*+} closure operators), the \kedavra{} count is lower than what we would report for an equivalent BNF-based \treevada{}/\toolName{} grammar. Without converting between EBNF and BNF, we can thus not directly compare \kedavra{} grammar sizes with \treevada{}/\toolName{} (which makes the findings presented in the \kedavra{} study hard to interpret).

\begin{table*}[h!t]
\centering
\caption{Inferred grammar statistics, 5-run average for \kedavra{}, skipping Crucio due to its low Table~\ref{table:main_table} F1 scores;
NT/T~=~(non-) terminals;
A~=~rules (alternatives);
l(A)~=~avg. rule length 
S~=~sum of rule lengths; %
D~=~longest minimum non-terminal distance from \emph{start};
Golden from Table~\ref{table:seed_stat};
\kedavra{}* undercounts \term{*}\term{+}\term{?} closures.
}
\label{table:stat_table}
\begin{tabular}{l|r r|r r r N{1}{1} r r| r r r N{1}{1} r r|r r r N{1}{1} r r}
\toprule
 &  \multicolumn{2}{c|}{Golden} & \multicolumn{6}{c|}{\treevada{}} & \multicolumn{6}{c|}{\kedavra{}} &  \multicolumn{6}{c}{\toolName{}} \\
\multicolumn{1}{c|}{} & \multicolumn{1}{c}{S} & \multicolumn{1}{c|}{D} & \multicolumn{1}{c}{NT} & \multicolumn{1}{c}{T} & \multicolumn{1}{c}{A} & \multicolumn{1}{c}{l(A)} & \multicolumn{1}{c}{S} & \multicolumn{1}{c|}{D} & \multicolumn{1}{c}{NT} & \multicolumn{1}{c}{T} & \multicolumn{1}{c}{A*} & \multicolumn{1}{c}{l(A)*} & \multicolumn{1}{c}{S*} & \multicolumn{1}{c|}{D} & \multicolumn{1}{c}{NT} & \multicolumn{1}{c}{T} & \multicolumn{1}{c}{A} & \multicolumn{1}{c}{l(A)} & \multicolumn{1}{c}{S} & \multicolumn{1}{c}{D} \\
\midrule

 \Use{turtle_programname} & \Use{turtle_grammar_NT_times_RHS} & \Use{turtle_grammar_ntd} & \Use{turtle_treevada_nt}  & \Use{turtle_treevada_T} & \Use{turtle_treevada_A} & \Use{turtle_treevada_IA} & \Use{turtle_treevada_S}  & \Use{turtle_treevada_D} &             \Use{turtle_Kedavra_nt}  & \Use{turtle_Kedavra_T} & \Use{turtle_Kedavra_A} & \Use{turtle_Kedavra_IA} & \Use{turtle_Kedavra_S}  & \Use{turtle_Kedavra_D} &             \Use{turtle_xvada_nt}  & \Use{turtle_xvada_T} & \Use{turtle_xvada_A} & \Use{turtle_xvada_IA} & \Use{turtle_xvada_S}  & \Use{turtle_xvada_D}  \\
 \Use{while_programname} & \Use{while_grammar_NT_times_RHS}  & \Use{while_grammar_ntd} & \Use{while_treevada_nt}  & \Use{while_treevada_T} & \Use{while_treevada_A} & \Use{while_treevada_IA} & \Use{while_treevada_S}  & \Use{while_treevada_D} &             \Use{while_Kedavra_nt}  & \Use{while_Kedavra_T} & \Use{while_Kedavra_A} & \Use{while_Kedavra_IA} & \Use{while_Kedavra_S}  & \Use{while_Kedavra_D} &             \Use{while_xvada_nt}  & \Use{while_xvada_T} & \Use{while_xvada_A} & \Use{while_xvada_IA} & \Use{while_xvada_S}  & \Use{while_xvada_D}  \\
 \Use{lisp_programname} & \Use{lisp_grammar_NT_times_RHS}  & \Use{lisp_grammar_ntd} & \Use{lisp_treevada_nt}  & \Use{lisp_treevada_T} & \Use{lisp_treevada_A} & \Use{lisp_treevada_IA} & \Use{lisp_treevada_S}  & \Use{lisp_treevada_D} &             \Use{lisp_Kedavra_nt}  & \Use{lisp_Kedavra_T} & \Use{lisp_Kedavra_A} & \Use{lisp_Kedavra_IA} & \Use{lisp_Kedavra_S}  & \Use{lisp_Kedavra_D} &             \Use{lisp_xvada_nt}  & \Use{lisp_xvada_T} & \Use{lisp_xvada_A} & \Use{lisp_xvada_IA} & \Use{lisp_xvada_S}  & \Use{lisp_xvada_D}  \\
 \Use{xml_programname} & \Use{xml_grammar_NT_times_RHS}  & \Use{xml_grammar_ntd} & \Use{xml_treevada_nt}  & \Use{xml_treevada_T} & \Use{xml_treevada_A} & \Use{xml_treevada_IA} & \Use{xml_treevada_S}  & \Use{xml_treevada_D} &             \Use{xml_Kedavra_nt}  & \Use{xml_Kedavra_T} & \Use{xml_Kedavra_A} & \Use{xml_Kedavra_IA} & \Use{xml_Kedavra_S}  & \Use{xml_Kedavra_D} &             \Use{xml_xvada_nt}  & \Use{xml_xvada_T} & \Use{xml_xvada_A} & \Use{xml_xvada_IA} & \Use{xml_xvada_S}  & \Use{xml_xvada_D}  \\
 \Use{json_programname} & \Use{json_grammar_NT_times_RHS}  & \Use{json_grammar_ntd} & \Use{json_treevada_nt}  & \Use{json_treevada_T} & \Use{json_treevada_A} & \Use{json_treevada_IA} & \Use{json_treevada_S}  & \Use{json_treevada_D} &             \Use{json_Kedavra_nt}  & \Use{json_Kedavra_T} & \Use{json_Kedavra_A} & \Use{json_Kedavra_IA} & \Use{json_Kedavra_S}  & \Use{json_Kedavra_D} &             \Use{json_xvada_nt}  & \Use{json_xvada_T} & \Use{json_xvada_A} & \Use{json_xvada_IA} & \Use{json_xvada_S}  & \Use{json_xvada_D}  \\
 \Use{tinyc_programname} & \Use{tinyc_grammar_NT_times_RHS}  & \Use{tinyc_grammar_ntd} & \Use{tinyc_treevada_nt}  & \Use{tinyc_treevada_T} & \Use{tinyc_treevada_A} & \Use{tinyc_treevada_IA} & \Use{tinyc_treevada_S}  & \Use{tinyc_treevada_D} &             \Use{tinyc_Kedavra_nt}  & \Use{tinyc_Kedavra_T} & \Use{tinyc_Kedavra_A} & \Use{tinyc_Kedavra_IA} & \Use{tinyc_Kedavra_S}  & \Use{tinyc_Kedavra_D} &             \Use{tinyc_xvada_nt}  & \Use{tinyc_xvada_T} & \Use{tinyc_xvada_A} & \Use{tinyc_xvada_IA} & \Use{tinyc_xvada_S}  & \Use{tinyc_xvada_D}  \\
 \Use{c-500_programname} & \Use{c-500_grammar_NT_times_RHS}  & \Use{tinyc_grammar_ntd} &  \Use{c-500_treevada_nt}  & \Use{c-500_treevada_T} & \Use{c-500_treevada_A} & \Use{c-500_treevada_IA} & \Use{c-500_treevada_S}  & \Use{c-500_treevada_D} &             \Use{c-500_Kedavra_nt}  & \Use{c-500_Kedavra_T} & \Use{c-500_Kedavra_A} & \Use{c-500_Kedavra_IA} & \Use{c-500_Kedavra_S}  & \Use{c-500_Kedavra_D} &             \Use{c-500_xvada_nt}  & \Use{c-500_xvada_T} & \Use{c-500_xvada_A} & \Use{c-500_xvada_IA} & \Use{c-500_xvada_S}  & \Use{c-500_xvada_D}  \\
 \Use{tiny_programname} & \Use{tiny_grammar_NT_times_RHS}  & \Use{tiny_grammar_ntd} &  \Use{tiny_treevada_nt}  & \Use{tiny_treevada_T} & \Use{tiny_treevada_A} & \Use{tiny_treevada_IA} & \Use{tiny_treevada_S}  & \Use{tiny_treevada_D} &             \Use{tiny_Kedavra_nt}  & \Use{tiny_Kedavra_T} & \Use{tiny_Kedavra_A} & \Use{tiny_Kedavra_IA} & \Use{tiny_Kedavra_S}  & \Use{tiny_Kedavra_D} &             \Use{tiny_xvada_nt}  & \Use{tiny_xvada_T} & \Use{tiny_xvada_A} & \Use{tiny_xvada_IA} & \Use{tiny_xvada_S}  & \Use{tiny_xvada_D}  \\
 \Use{minic_programname} & \Use{minic_grammar_NT_times_RHS}  & \Use{minic_grammar_ntd} &  \Use{minic_treevada_nt}  & \Use{minic_treevada_T} & \Use{minic_treevada_A} & \Use{minic_treevada_IA} & \Use{minic_treevada_S}  & \Use{minic_treevada_D} &             \Use{minic_Kedavra_nt}  & \Use{minic_Kedavra_T} & \Use{minic_Kedavra_A} & \Use{minic_Kedavra_IA} & \Use{minic_Kedavra_S}  & \Use{minic_Kedavra_D} &             \Use{minic_xvada_nt}  & \Use{minic_xvada_T} & \Use{minic_xvada_A} & \Use{minic_xvada_IA} & \Use{minic_xvada_S}  & \Use{curl_xvada_D}  \\
 \Use{curl_programname} & \Use{curl_grammar_NT_times_RHS}  & \Use{curl_grammar_ntd} &  \Use{curl_treevada_nt}  & \Use{curl_treevada_T} & \Use{curl_treevada_A} & \Use{curl_treevada_IA} & \Use{curl_treevada_S}  & \Use{curl_treevada_D} &             \Use{curl_Kedavra_nt}  & \Use{curl_Kedavra_T} & \Use{curl_Kedavra_A} & \Use{curl_Kedavra_IA} & \Use{curl_Kedavra_S}  & \Use{curl_Kedavra_D} &             \Use{curl_xvada_nt}  & \Use{curl_xvada_T} & \Use{curl_xvada_A} & \Use{curl_xvada_IA} & \Use{curl_xvada_S}  & \Use{curl_xvada_D}  \\
\midrule
 \Use{Avg_lo_programname} & \Use{Avg-Small-Program_grammar_NT_times_RHS} & \Use{Avg-Small-Program_grammar_ntd} &  \Use{Avg_lo_treevada_nt}  & \Use{Avg_lo_treevada_T} & \Use{Avg_lo_treevada_A} & \Use{Avg_lo_treevada_IA} & \Use{Avg_lo_treevada_S}  & \Use{Avg_lo_treevada_D} &             \Use{Avg_lo_Kedavra_nt}  & \Use{Avg_lo_Kedavra_T} & \Use{Avg_lo_Kedavra_A} & \Use{Avg_lo_Kedavra_IA} & \Use{Avg_lo_Kedavra_S}  & \Use{Avg_lo_Kedavra_D} &             \Use{Avg_lo_xvada_nt}  & \Use{Avg_lo_xvada_T} & \Use{Avg_lo_xvada_A} & \Use{Avg_lo_xvada_IA} & \Use{Avg_lo_xvada_S}  & \Use{Avg_lo_xvada_D}  \\
\midrule
\Use{liquid_programname} & n/a & \Use{liquid_grammar_ntd} &  \Use{liquid_treevada_nt}  & \Use{liquid_treevada_T} & \Use{liquid_treevada_A} & \Use{liquid_treevada_IA} & \Use{liquid_treevada_S}  & \Use{liquid_treevada_D} &             \multicolumn{6}{c|}{n/a} &             \Use{liquid_xvada_nt}  & \Use{liquid_xvada_T} & \Use{liquid_xvada_A} & \Use{liquid_xvada_IA} & \Use{liquid_xvada_S}  & \Use{liquid_xvada_D}  \\
 \Use{lua_programname} & \Use{lua_grammar_NT_times_RHS} & \Use{lua_grammar_ntd} &  \Use{lua_treevada_nt}  & \Use{lua_treevada_T} & \Use{lua_treevada_A} & \Use{lua_treevada_IA} & \Use{lua_treevada_S}  & \Use{lua_treevada_D} &             \multicolumn{6}{c|}{n/a} &             \Use{lua_xvada_nt}  & \Use{lua_xvada_T} & \Use{lua_xvada_A} & \Use{lua_xvada_IA} & \Use{lua_xvada_S}  & \Use{lua_xvada_D}  \\
 \Use{c_programname} & \Use{c_grammar_NT_times_RHS} & \Use{c_grammar_ntd} &  \Use{c_treevada_nt}  & \Use{c_treevada_T} & \Use{c_treevada_A} & \Use{c_treevada_IA} & \Use{c_treevada_S}  & \Use{c_treevada_D} &  \multicolumn{6}{c|}{n/a} &             \Use{c_xvada_nt}  & \Use{c_xvada_T} & \Use{c_xvada_A} & \Use{c_xvada_IA} & \Use{c_xvada_S}  & \Use{c_xvada_D}  \\
 \Use{java_programname} & \Use{java_grammar_NT_times_RHS}  & \Use{java_grammar_ntd} &  \Use{java_treevada_nt}  & \Use{java_treevada_T} & \Use{java_treevada_A} & \Use{java_treevada_IA} & \Use{java_treevada_S}  & \Use{java_treevada_D} & \multicolumn{6}{c|}{n/a} &             \Use{java_xvada_nt}  & \Use{java_xvada_T} & \Use{java_xvada_A} & \Use{java_xvada_IA} & \Use{java_xvada_S}  & \Use{java_xvada_D}  \\
 \Use{cpp_programname} & \Use{cpp_grammar_NT_times_RHS} & \Use{cpp_grammar_ntd} &  \Use{cpp_treevada_nt}  & \Use{cpp_treevada_T} & \Use{cpp_treevada_A} & \Use{cpp_treevada_IA} & \Use{cpp_treevada_S}  & \Use{cpp_treevada_D} &   \multicolumn{6}{c|}{n/a} &             \Use{cpp_xvada_nt}  & \Use{cpp_xvada_T} & \Use{cpp_xvada_A} & \Use{cpp_xvada_IA} & \Use{cpp_xvada_S}  & \Use{cpp_xvada_D}  \\
 \Use{rust_programname} & \Use{rust_grammar_NT_times_RHS} & \Use{rust_grammar_ntd} &  \Use{rust_treevada_nt}  & \Use{rust_treevada_T} & \Use{rust_treevada_A} & \Use{rust_treevada_IA} & \Use{rust_treevada_S}  & \Use{rust_treevada_D} &  \multicolumn{6}{c|}{n/a} &             \Use{rust_xvada_nt}  & \Use{rust_xvada_T} & \Use{rust_xvada_A} & \Use{rust_xvada_IA} & \Use{rust_xvada_S}  & \Use{rust_xvada_D}  \\
 \Use{mysql_programname} & \Use{mysql_grammar_NT_times_RHS} & \Use{mysql_grammar_ntd} &  \Use{mysql_treevada_nt}  & \Use{mysql_treevada_T} & \Use{mysql_treevada_A} & \Use{mysql_treevada_IA} & \Use{mysql_treevada_S}  & \Use{mysql_treevada_D} &  \multicolumn{6}{c|}{n/a} &             \Use{mysql_xvada_nt}  & \Use{mysql_xvada_T} & \Use{mysql_xvada_A} & \Use{mysql_xvada_IA} & \Use{mysql_xvada_S}  & \Use{mysql_xvada_D}  \\
\midrule
 \Use{Avg_hi_programname} & \Use{Avg-Big-Program_grammar_NT_times_RHS} & \Use{Avg-Big-Program_grammar_ntd} & \Use{Avg_hi_treevada_nt}  & \Use{Avg_hi_treevada_T} & \Use{Avg_hi_treevada_A} & \Use{Avg_hi_treevada_IA} & \Use{Avg_hi_treevada_S}  & \Use{Avg_hi_treevada_D} &  \multicolumn{6}{c|}{n/a} &             \Use{Avg_hi_xvada_nt}  & \Use{Avg_hi_xvada_T} & \Use{Avg_hi_xvada_A} & \Use{Avg_hi_xvada_IA} & \Use{Avg_hi_xvada_S}  & \Use{Avg_hi_xvada_D}  \\
\bottomrule
\end{tabular}
\end{table*}

Table~\ref{table:stat_table} column \textit{S} shows that \Tool{} generally infers more compact grammars than \treevada{}. On the Lo subjects, \Tool{} reduces average $S$ from \Use{Avg_lo_treevada_S} to \Use{Avg_lo_xvada_S} and is smaller than \treevada{} on 8 of 10 languages. The reduction is especially visible for \textit{turtle}, \textit{json}, and \textit{minic}. On the Hi subjects, the difference is larger: average $S$ decreases from \Use{Avg_hi_treevada_S} to \Use{Avg_hi_xvada_S}, and \Tool{} is smaller on 5 of 7 languages, especially for \textit{c}, \textit{java}, \textit{cpp}, and \textit{rust}.

\begin{framed}
\noindent
\emph{
\textbf{\RQCompact{}: }
While on average \Tool{} is more accurate than \treevada{} (Table~\ref{table:main_table}), \Tool{}-inferred grammars on average are also more compact (Table~\ref{table:stat_table}), reducing average grammar size by 7\% (Lo) and 20\% (Hi).
}
\end{framed}

\subsection{\RQAblation{}: HDD's Contribution to Accuracy}

\begin{table}[h!t]
\centering
\caption{Ablation: \toolName{} F1 scores with \& w/o HDD on random 5-seed (Lo) and 10-seed (Hi) inputs.
}
\label{table:hdd_table}
\resizebox{\columnwidth}{!}{%
\begin{tabular}{l|r r r| r r r|r|r}
\toprule
 &  \multicolumn{3}{c|}{\toolName{} w/o HDD} &  \multicolumn{3}{c|}{\toolName{}} & \multicolumn{1}{c|}{HDD} & \multicolumn{1}{c}{Total} \\
\multicolumn{1}{c|}{}  & \multicolumn{1}{c}{p} & \multicolumn{1}{c}{r} & \multicolumn{1}{c|}{f1} & \multicolumn{1}{c}{p} & \multicolumn{1}{c}{r} & \multicolumn{1}{c|}{f1} & \multicolumn{1}{c|}{t[s]} & \multicolumn{1}{c}{t[s]} \\
\midrule
\Use{turtle_programname} & \Use{turtle_xvada_precision_without_hdd}  & \Use{turtle_xvada_recall_without_hdd} & \Use{turtle_xvada_fscore_without_hdd} & \Use{turtle_xvada_precision_with_hdd} & \Use{turtle_xvada_recall_with_hdd}  & \Use{turtle_xvada_fscore_with_hdd} & \Use{turtle_xvada_hdd_time}  & \Use{turtle_xvada_total_time} \\
\Use{while_programname} & \Use{while_xvada_precision_without_hdd}  & \Use{while_xvada_recall_without_hdd} & \Use{while_xvada_fscore_without_hdd} & \Use{while_xvada_precision_with_hdd} & \Use{while_xvada_recall_with_hdd}  & \Use{while_xvada_fscore_with_hdd} & \Use{while_xvada_hdd_time}  & \Use{while_xvada_total_time} \\
\Use{lisp_programname} & \Use{lisp_xvada_precision_without_hdd}  & \Use{lisp_xvada_recall_without_hdd} & \Use{lisp_xvada_fscore_without_hdd} & \Use{lisp_xvada_precision_with_hdd} & \Use{lisp_xvada_recall_with_hdd}  & \Use{lisp_xvada_fscore_with_hdd} & \Use{lisp_xvada_hdd_time}  & \Use{lisp_xvada_total_time} \\
\Use{xml_programname} & \Use{xml_xvada_precision_without_hdd}  & \Use{xml_xvada_recall_without_hdd} & \Use{xml_xvada_fscore_without_hdd} & \Use{xml_xvada_precision_with_hdd} & \Use{xml_xvada_recall_with_hdd}  & \Use{xml_xvada_fscore_with_hdd} & \Use{xml_xvada_hdd_time}  & \Use{xml_xvada_total_time} \\
\Use{json_programname} & \Use{json_xvada_precision_without_hdd}  & \Use{json_xvada_recall_without_hdd} & \Use{json_xvada_fscore_without_hdd} & \Use{json_xvada_precision_with_hdd} & \Use{json_xvada_recall_with_hdd}  & \Use{json_xvada_fscore_with_hdd} & \Use{json_xvada_hdd_time}  & \Use{json_xvada_total_time} \\
\Use{tinyc_programname} & \Use{tinyc_xvada_precision_without_hdd}  & \Use{tinyc_xvada_recall_without_hdd} & \Use{tinyc_xvada_fscore_without_hdd} & \Use{tinyc_xvada_precision_with_hdd} & \Use{tinyc_xvada_recall_with_hdd}  & \Use{tinyc_xvada_fscore_with_hdd} & \Use{tinyc_xvada_hdd_time}  & \Use{tinyc_xvada_total_time} \\
\Use{c-500_programname} & \Use{c-500_xvada_precision_without_hdd}  & \Use{c-500_xvada_recall_without_hdd} & \Use{c-500_xvada_fscore_without_hdd} & \Use{c-500_xvada_precision_with_hdd} & \Use{c-500_xvada_recall_with_hdd}  & \Use{c-500_xvada_fscore_with_hdd} & \Use{c-500_xvada_hdd_time}  & \Use{c-500_xvada_total_time} \\
\Use{tiny_programname} & \Use{tiny_xvada_precision_without_hdd}  & \Use{tiny_xvada_recall_without_hdd} & \Use{tiny_xvada_fscore_without_hdd} & \Use{tiny_xvada_precision_with_hdd} & \Use{tiny_xvada_recall_with_hdd}  & \Use{tiny_xvada_fscore_with_hdd} & \Use{tiny_xvada_hdd_time}  & \Use{tiny_xvada_total_time} \\
\Use{minic_programname} & \Use{minic_xvada_precision_without_hdd}  & \Use{minic_xvada_recall_without_hdd} & \Use{minic_xvada_fscore_without_hdd} & \Use{minic_xvada_precision_with_hdd} & \Use{minic_xvada_recall_with_hdd}  & \Use{minic_xvada_fscore_with_hdd} & \Use{minic_xvada_hdd_time}  & \Use{minic_xvada_total_time} \\
\Use{curl_programname} & \Use{curl_xvada_precision_without_hdd}  & \Use{curl_xvada_recall_without_hdd} & \Use{curl_xvada_fscore_without_hdd} & \Use{curl_xvada_precision_with_hdd} & \Use{curl_xvada_recall_with_hdd}  & \Use{curl_xvada_fscore_with_hdd} & \Use{curl_xvada_hdd_time}  & \Use{curl_xvada_total_time} \\
\midrule
\Use{Avg_lo_programname} & \Use{Avg_lo_xvada_precision_without_hdd}  & \Use{Avg_lo_xvada_recall_without_hdd} & \Use{Avg_lo_xvada_fscore_without_hdd} & \Use{Avg_lo_xvada_precision_with_hdd} & \Use{Avg_lo_xvada_recall_with_hdd}  & \Use{Avg_lo_xvada_fscore_with_hdd} & \Use{Avg_lo_xvada_hdd_time}  & \Use{Avg_lo_xvada_total_time} \\
\midrule
\Use{liquid_programname} & \Use{liquid_xvada_precision_without_hdd}  & \Use{liquid_xvada_recall_without_hdd} & \Use{liquid_xvada_fscore_without_hdd} & \Use{liquid_xvada_precision_with_hdd} & \Use{liquid_xvada_recall_with_hdd}  & \Use{liquid_xvada_fscore_with_hdd} & \Use{liquid_xvada_hdd_time}  & \Use{liquid_xvada_total_time} \\
\Use{lua_programname} & \Use{lua_xvada_precision_without_hdd}  & \Use{lua_xvada_recall_without_hdd} & \Use{lua_xvada_fscore_without_hdd} & \Use{lua_xvada_precision_with_hdd} & \Use{lua_xvada_recall_with_hdd}  & \Use{lua_xvada_fscore_with_hdd} & \Use{lua_xvada_hdd_time}  & \Use{lua_xvada_total_time} \\
\Use{c_programname} & \Use{c_xvada_precision_without_hdd}  & \Use{c_xvada_recall_without_hdd} & \Use{c_xvada_fscore_without_hdd} & \Use{c_xvada_precision_with_hdd} & \Use{c_xvada_recall_with_hdd}  & \Use{c_xvada_fscore_with_hdd} & \Use{c_xvada_hdd_time}  & \Use{c_xvada_total_time} \\
\Use{java_programname} & \Use{java_xvada_precision_without_hdd}  & \Use{java_xvada_recall_without_hdd} & \Use{java_xvada_fscore_without_hdd} & \Use{java_xvada_precision_with_hdd} & \Use{java_xvada_recall_with_hdd}  & \Use{java_xvada_fscore_with_hdd} & \Use{java_xvada_hdd_time}  & \Use{java_xvada_total_time} \\
\Use{cpp_programname} & \Use{cpp_xvada_precision_without_hdd}  & \Use{cpp_xvada_recall_without_hdd} & \Use{cpp_xvada_fscore_without_hdd} & \Use{cpp_xvada_precision_with_hdd} & \Use{cpp_xvada_recall_with_hdd}  & \Use{cpp_xvada_fscore_with_hdd} & \Use{cpp_xvada_hdd_time}  & \Use{cpp_xvada_total_time} \\
\Use{rust_programname} & \Use{rust_xvada_precision_without_hdd}  & \Use{rust_xvada_recall_without_hdd} & \Use{rust_xvada_fscore_without_hdd} & \Use{rust_xvada_precision_with_hdd} & \Use{rust_xvada_recall_with_hdd}  & \Use{rust_xvada_fscore_with_hdd} & \Use{rust_xvada_hdd_time}  & \Use{rust_xvada_total_time} \\
\Use{mysql_programname} & \Use{mysql_xvada_precision_without_hdd}  & \Use{mysql_xvada_recall_without_hdd} & \Use{mysql_xvada_fscore_without_hdd} & \Use{mysql_xvada_precision_with_hdd} & \Use{mysql_xvada_recall_with_hdd}  & \Use{mysql_xvada_fscore_with_hdd} & \Use{mysql_xvada_hdd_time}  & \Use{mysql_xvada_total_time} \\
\midrule
\Use{Avg_hi_programname} & \Use{Avg_hi_xvada_precision_without_hdd}  & \Use{Avg_hi_xvada_recall_without_hdd} & \Use{Avg_hi_xvada_fscore_without_hdd} & \Use{Avg_hi_xvada_precision_with_hdd} & \Use{Avg_hi_xvada_recall_with_hdd}  & \Use{Avg_hi_xvada_fscore_with_hdd} & \Use{Avg_hi_xvada_hdd_time}  & \Use{Avg_hi_xvada_total_time} \\
\bottomrule
\end{tabular}
} %
\end{table}

Table~\ref{table:hdd_table} isolates HDD's contribution by comparing \toolName{} with and without HDD. To expose HDD's impact, we want large seed programs to be less likely to exhibit simple grammar-rule alternatives. We thus use a random seed subset (5 of the Table~\ref{table:seed_stat} seed programs for Lo, 10 for Hi).

The main effect of HDD is improved recall (on average from 49 to~52\% on Lo, raising average F1 from 55 to~58\%, with average precision unchanged at 85\%). On Hi the effect is larger (except rust): average recall increases from 25 to 29\%, average precision from 55 to~60\%, and average F1 from 26 to~32\%. The largest gains occur when the reduced seed sets are mostly compositions of smaller complete program fragments. For example, liquid improves from 42 to~68\% F1, c from 29 to~43\%, and lua from 2 to~9\%. In contrast, subjects whose reduced seed set already exhibits the relevant alternatives show little or no F1 change. Only in rust, F1 drops indicating HDD added incorrect decomposed alternatives. For both language groups the average HDD runtime overhead is below 1\%, i.e., 1.4 of 206 seconds (Lo) and 20 of 7,912 seconds (Hi).

\begin{framed}
\noindent
\emph{
\textbf{\RQAblation{}: }
With an average sub-1\% runtime overhead on our reduced-seed inputs, HDD increased the average F1 socre by 3pp (Lo) and 6pp (Hi).
}
\end{framed}

\subsection{\RQFuzzing{}: Fuzzing the Python Liquid Parser: 5 Bugs \& a CVE}

\Tool{} learns from a black-box parser and encodes the parser's accept/reject behavior in the inferred grammar. This allows a grammar-based fuzzer to mutate parser-specific derivation trees, helping it reach deeply nested parser states.

Liquid is Shopify's template language, first implemented in Ruby and since adopted widely (with implementations in python, js, c\#, etc.). For \RQFuzzing{} we pass the \Tool{}-inferred Liquid grammar to the coverage-driven \& grammar-aware Nautilus fuzzer~\cite{aschermann2019nautilus} to find bugs in the python Liquid engine, which (according to PyPIStats) gets over 1M monthly downloads\footnote{\Accessed \url{https://pypistats.org/packages/python-liquid}}. (\toolName{}'s grammar inference is blackbox and does not use coverage data; only the fuzzer is greybox.) As the public Nautilus artifact is no longer maintained, we use LibAFL~\cite{fioraldi2022libafl}. The fuzzer requires grammars to include whitespace explicitly, which \toolName{} supports. We thus provide the inferred production rules directly to the fuzzer in json format.

\MyPara{Findings}
The highest-impact finding was the denial-of-service (DoS) vulnerability CVE-2026-55865 \toolName{} found during black-box grammar inference (without fuzzing). A malformed \term{case} tag with no terminating \term{endcase} yielded an infinite loop in Python Liquid. An attacker who can submit templates could thus prevent parsing from terminating and consume service resources.

Fuzzing then found five additional bugs (all since fixed by Python Liquid developers). These bugs involve improperly handling malformed or missing tag arguments, nested bracket unfolding, and nested markup blocks, leading to resource-exhaustion failures or uncaught runtime exceptions such as \texttt{RecursionError} and \texttt{MemoryError}.

\begin{framed}
\noindent
\emph{
\textbf{\RQFuzzing{}: }
\Tool{} itself found a DoS CVE in Python Liquid. Directly using the \Tool{}-inferred grammar, the Nautilus fuzzer found five more since-fixed Python Liquid bugs.
}
\end{framed}

\section{Threats to Validity}
\label{sec:threats}

Having a \textit{representative sample}~\cite{angluin1981note} (a finite set of positive examples where all grammar rules are exercised) is crucial for inferred grammar's coverage. To best approximate this condition, we sample from golden grammar (where possible) using a standalone fuzzer instead of using real programs as seed/test sets. Real programs tend to be biased toward commonly used parts of a language~\cite{aschermann2019nautilus}, and therefore may leave many grammar rules unexercised. While this choice improves language coverage, but may make our seed/test sets differ from naturally occurring programs.

\section{Related Work}
\label{sec:related}

Large language models (LLMs) have made significant advancements and show remarkable ability in many software engineering tasks. These models demonstrate robust performance in code understanding, enabling effective code summarization, refactoring, verification, and auto-completion. However, theoretical~\cite{Hahnlimitation} and empirical studies~\cite{bhattamishra2020ability, ebrahimi-etal-2020-self, delétang2023neuralnetworkschomskyhierarchy, zhang2024transformer} suggest that LLMs have limited ability to learn formal grammars (such as context-free grammars) from sample programs.

Grammar inference is a long-studied problem. Angluin's seminal \textit{active learning} work shows how to learn regular languages from queries and counterexamples~\cite{angluin1987learning}, and also discusses context-free learning under stronger assumptions~\cite{angluin1988queries}. For program inputs, however, these assumptions rarely hold: a parser typically provides only membership queries, and query complexity grows quickly beyond regular languages~\cite{querycomplexity}. \glade{} showed that this black-box setting can still be effective by learning seed-specific structure and using oracle feedback to generalize input grammars~\cite{bastani2017synthesizing}.

A second line of work connects grammar inference with compression. Sequitur replaces repeated adjacent sequences with hierarchical rules~\cite{sequitur1997}. The already discussed \arvada{}, \treevada{}, \kedavra{}, and \crucio{} adapt this intuition to black-box context-free inference.

Besides oracle answers, white-box approaches also leverage parser internals. \textsc{Autogram} mines grammars from dynamic taints~\cite{hoschele2016mining}, while \textsc{Mimid} maps dynamic control-flow contexts to grammar structure~\cite{gopinath2020mining}. These techniques can produce well-labeled grammars, but require access to parser internals. More recent static approaches reduce the dependence on seed inputs: \textsc{Panini} infers regular grammars for short ad hoc parsers using summaries of string operations~\cite{schroder2025static}, and \textsc{Stalagmite} uses symbolic parsing to infer context-free grammars from recursive-descent parser code~\cite{bettscheider2025inferring}. 
These approaches are complementary to black-box grammar inference: they exploit source code when available, whereas our setting assumes only sample programs and a black-box parser oracle.

\section{Conclusions}
\label{sec:conclusion}

\Tool{} introduced several new techniques for deterministic inference of context-free grammars. In an empirical comparison that avoids several pitfalls of recent studies, \Tool{} improved on the highest-scoring competitor (\treevada{}) both in grammar accuracy and grammar compactness. \toolName{} also found a CVE in the widely used Python Liquid engine. Fuzzing based on the \toolName{}-inferred grammar found five more bugs, which the Python Liquid developers fixed based on our bug reports. \toolName{} and all experimental data and scripts are publicly available.

\bibliographystyle{IEEEtran}
\bibliography{ref}

@article{Hahnlimitation,
    author = {Hahn, Michael},
    title = {Theoretical Limitations of Self-Attention in Neural Sequence Models},
    journal = {Transactions of the Association for Computational Linguistics},
    volume = {8},
    pages = {156-171},
    year = {2020},
    month = {01},
    doi = {10.1162/tacl_a_00306},
}

@inproceedings{bhattamishra2020ability,
  author       = {Satwik Bhattamishra and
                  Kabir Ahuja and
                  Navin Goyal},
  title={On the ability and limitations of transformers to recognize formal languages},
  booktitle    = {Proc. Conference on Empirical Methods in Natural
                  Language Processing (EMNLP)},
  pages        = {7096--7116},
  publisher    = {Association for Computational Linguistics},
  year         = {2020},
  month = nov,
  url          = {https://doi.org/10.18653/v1/2020.emnlp-main.576},
  doi          = {10.18653/V1/2020.EMNLP-MAIN.576},
}

@inproceedings{ebrahimi-etal-2020-self,
    title = "How Can Self-Attention Networks Recognize {D}yck-n Languages?",
    author = "Ebrahimi, Javid  and
      Gelda, Dhruv  and
      Zhang, Wei",
    booktitle = "Findings of the Association for Computational Linguistics: EMNLP 2020",
    month = nov,
    year = "2020",
    publisher = "Association for Computational Linguistics",
    url = "https://aclanthology.org/2020.findings-emnlp.384/",
    doi = "10.18653/v1/2020.findings-emnlp.384",
    pages = "4301--4306"
}

@inproceedings{delétang2023neuralnetworkschomskyhierarchy,
  author       = {Gr{\'{e}}goire Del{\'{e}}tang and
                  Anian Ruoss and
                  Jordi Grau{-}Moya and
                  Tim Genewein and
                  Li Kevin Wenliang and
                  Elliot Catt and
                  Chris Cundy and
                  Marcus Hutter and
                  Shane Legg and
                  Joel Veness and
                  Pedro A. Ortega},
  title        = {Neural Networks and the {Chomsky} Hierarchy},
  booktitle    = {Proc. 11th International Conference on Learning Representations (ICLR)},
  year         = {2023},
  url={https://openreview.net/forum?id=WbxHAzkeQcn},
}

@inproceedings{kulkarni2021learning,
  title={Learning Highly Recursive Input Grammars},
  author={Kulkarni, Neil and Lemieux, Caroline and Sen, Koushik},
  booktitle={Proc. 36th IEEE/ACM International Conference on Automated Software Engineering (ASE)},
  pages={456--467},
  year={2021},
  publisher={IEEE}
}

@inproceedings{arefin2024fast,
  title={Fast deterministic black-box context-free grammar inference},
  author={Arefin, Mohammad Rifat and Shetiya, Suraj and Wang, Zili and Csallner, Christoph},
  booktitle={Proc. IEEE/ACM 46th International Conference on Software Engineering (ICSE)},
  pages={1--12},
  year={2024}
}

@inproceedings{bastani2017synthesizing,
  author    = {Osbert Bastani and
               Rahul Sharma and
               Alex Aiken and
               Percy Liang},
  title     = {Synthesizing program input grammars},
  booktitle = {Proc. 38th {ACM} {SIGPLAN} Conference on Programming
               Language Design and Implementation (PLDI)},
  pages     = {95--110},
  publisher = {{ACM}},
  year      = {2017},
  month = jun,
}

@inproceedings{li2024incremental,
  title={Incremental Context-free Grammar Inference in Black Box Settings},
  author={Li, Feifei and Chen, Xiao and Xiao, Xi and Sun, Xiaoyu and Chen, Chuan and Wang, Shaohua and Han, Jitao},
  booktitle={Proc. 39th IEEE/ACM International Conference on Automated Software Engineering (ASE)},
  pages={1171--1182},
  year={2024}
}

@InProceedings{querycomplexity,
author="Domingo, Carlos
and Lav{\'i}n, V{\'i}ctor",
title="The query complexity of learning some subclasses of context-free grammars",
booktitle="Computational Learning Theory",
year="1995",
publisher="Springer",
pages="404--414",
}

@ARTICLE{hddzeller,
  author={Zeller, A. and Hildebrandt, R.},
  journal={IEEE Transactions on Software Engineering}, 
  title={Simplifying and isolating failure-inducing input}, 
  year={2002},
  volume={28},
  number={2},
  pages={183--200},
  doi={10.1109/32.988498}}

@inproceedings{hdd,
author = {Misherghi, Ghassan and Su, Zhendong},
title = {{HDD}: Hierarchical delta debugging},
year = {2006},
publisher = {ACM},
url = {https://doi.org/10.1145/1134285.1134307},
doi = {10.1145/1134285.1134307},
booktitle = {Proc. 28th International Conference on Software Engineering (ICSE)},
pages = {142--151},
}

@article{sequitur1997,
  author       = {Craig G. Nevill{-}Manning and
                  Ian H. Witten},
  title        = {Identifying Hierarchical Structure in Sequences: {A} linear-time algorithm},
  journal      = {Journal of Artificial Intelligence Research},
  volume       = {7},
  pages        = {67--82},
  year         = {1997},
  url          = {https://doi.org/10.1613/jair.374},
  doi          = {10.1613/JAIR.374},
}

@inproceedings{bendrissou2022synthesizing,
  title={``{Synthesizing} input grammars'': A replication study},
  author={Bachir Bendrissou and
                  Rahul Gopinath and
                  Andreas Zeller},
  booktitle={Proc. 43rd ACM SIGPLAN International Conference on Programming Language Design and Implementation (PLDI)},
  pages={260--268},
  year={2022}
}

@misc{bachir_bendrissou_artifact,
  author       = {Bachir Bendrissou and
                  Rahul Gopinath and
                  Andreas Zeller},
  title        = {Replication package for ``{Synthesizing} Input
                   Grammars: A Replication Study''},
  month        = mar,
  year         = 2022,
  publisher    = {Zenodo},
  version      = {0.3},
  doi          = {10.5281/zenodo.6460021},
  url          = {https://doi.org/10.5281/zenodo.6460021},
}

@inproceedings{gopinath2020mining,
  title={Mining input grammars from dynamic control flow},
  author={Gopinath, Rahul and Mathis, Bj{\"o}rn and Zeller, Andreas},
  booktitle={Proc. 28th ACM joint meeting on european software engineering conference and symposium on the foundations of software engineering},
  pages={172--183},
  year={2020}
}

@article{schroder2025static,
  title={Static Inference of Regular Grammars for Ad Hoc Parsers},
  author={Schr{\"o}der, Michael and Cito, J{\"u}rgen},
  journal={Proc. ACM on Programming Languages},
  volume={9},
  number={OOPSLA2},
  pages={113--143},
  year={2025},
  publisher={ACM}
}

@article{bettscheider2025inferring,
  title={Inferring Input Grammars from Code with Symbolic Parsing},
  author={Bettscheider, Leon and Zeller, Andreas},
  journal={ACM Transactions on Software Engineering and Methodology},
  year={2025},
  publisher={ACM}
}

@inproceedings{hodovan2018grammarinator,
author = {Hodov\'{a}n, Ren\'{a}ta and Kiss, \'{A}kos and Gyim\'{o}thy, Tibor},
title = {Grammarinator: a grammar-based open source fuzzer},
year = {2018},
isbn = {9781450360531},
publisher = {ACM},
url = {https://doi.org/10.1145/3278186.3278193},
doi = {10.1145/3278186.3278193},
booktitle = {Proc. 9th ACM SIGSOFT International Workshop on Automating TEST Case Design, Selection, and Evaluation},
pages = {45–48},
numpages = {4},
series = {A-TEST 2018}
}

@article{fandango2025amaya,
author = {Zamudio Amaya, Jos\'{e} Antonio and Smytzek, Marius and Zeller, Andreas},
title = {FANDANGO: Evolving Language-Based Testing},
year = {2025},
issue_date = {July 2025},
publisher = {ACM},
volume = {2},
number = {ISSTA},
url = {https://doi.org/10.1145/3728915},
doi = {10.1145/3728915},
journal = {Proc. ACM Softw. Eng.},
month = jun,
articleno = {ISSTA040},
numpages = {23}
}

@misc{li2026contextfreegrammarinferencecomplex,
      title={Context-Free Grammar Inference for Complex Programming Languages in Black Box Settings}, 
      author={Feifei Li and Xiao Chen and Xiaoyu Sun and Xi Xiao and Shaohua Wang and Yong Ding and Sheng Wen and Qing Li},
      year={2026},
      eprint={2601.12385},
      archivePrefix={arXiv},
      primaryClass={cs.PL},
      url={https://arxiv.org/abs/2601.12385}, 
}

@article{Arefin2023,
author = "Mohammad Rifat Arefin and Suraj Shetiya and Zili Wang and Christoph Csallner",
title = "{Artifact for Fast Deterministic Black-box Context-free Grammar Inference}",
year = "2023",
month = "12",
url = "https://figshare.com/articles/conference_contribution/Fast_Deterministic_Black-box_Context-free_Grammar_Inference/23907738",
doi = "10.6084/m9.figshare.23907738.v6"
}

@article{zhang2024transformer,
  author       = {Shizhuo Dylan Zhang and
                  Curt Tigges and
                  Zory Zhang and
                  Stella Biderman and
                  Maxim Raginsky and
                  Talia Ringer},
  title        = {Transformer-based models are not yet perfect at learning to emulate
                  structural recursion},
  journal      = {Transactions on Machine Learning Research},
  volume       = {2024},
  year         = {2024},
  url          = {https://openreview.net/forum?id=Ry5CXXm1sf},
}

@inproceedings{van2019lightweight,
  title={Lightweight multi-language syntax transformation with parser parser combinators},
  author={Van Tonder, Rijnard and Le Goues, Claire},
  booktitle={Proc. 40th ACM SIGPLAN Conference on Programming Language Design and Implementation},
  pages={363--378},
  year={2019}
}

@article{Crepinsek2010,
  author       = {Matej Crepinsek and
                  Tomaz Kosar and
                  Marjan Mernik and
                  Julien Cervelle and
                  R{\'{e}}mi Forax and
                  Gilles Roussel},
  title        = {On automata and language based grammar metrics},
  journal      = {Computer Science and Information Systems},
  volume       = {7},
  number       = {2},
  pages        = {309--329},
  year         = {2010},
  doi          = {10.2298/CSIS1002309C},
}

@article{Power2004,
  author       = {James F. Power and
                  Brian A. Malloy},
  title        = {A metrics suite for grammar-based software},
  journal      = { Journal of Software Maintenance and Evolution: Research and Practice},
  volume       = {16},
  number       = {6},
  pages        = {405--426},
  year         = {2004},
  doi          = {10.1002/SMR.293},
}

@article{Csuhaj-Varju1993,
  author       = {Erzs{\'{e}}bet Csuhaj{-}Varj{\'{u}} and
                  Alica Kelemenov{\'{a}}},
  title        = {Descriptional complexity of context-free grammar forms},
  journal      = {Theoretical Computer Science},
  volume       = {112},
  number       = {2},
  pages        = {277--289},
  year         = {1993},
  doi          = {10.1016/0304-3975(93)90021-K},
}

@book{Aho2006,
author = {Aho, Alfred V. and Lam, Monica S. and Sethi, Ravi and Ullman, Jeffrey D.},
title = {Compilers: Principles, Techniques, and Tools},
year = {2006},
isbn = {0321486811},
publisher = {Addison-Wesley},
edition = {2nd}
}

@inproceedings{fioraldi2022libafl,
  title={{LibAFL}: A framework to build modular and reusable fuzzers},
  author={Fioraldi, Andrea and Maier, Dominik Christian and Zhang, Dongjia and Balzarotti, Davide},
  booktitle={Proc. 2022 ACM SIGSAC Conference on Computer and Communications Security},
  pages={1051--1065},
  year={2022}
}

@inproceedings{aschermann2019nautilus,
  title={NAUTILUS: Fishing for deep bugs with grammars.},
  author={Aschermann, Cornelius and Frassetto, Tommaso and Holz, Thorsten and Jauernig, Patrick and Sadeghi, Ahmad-Reza and Teuchert, Daniel},
  booktitle={NDSS},
  volume={19},
  pages={337},
  year={2019}
}

@article{sakakibara1992efficient,
  title={Efficient learning of context-free grammars from positive structural examples},
  author={Sakakibara, Yasubumi},
  journal={Information and Computation},
  volume={97},
  number={1},
  pages={23--60},
  year={1992},
  publisher={Elsevier}
}

@book{reghizzi2013formal,
  title={Formal languages and compilation},
  author={Reghizzi, Stefano Crespi and Breveglieri, Luca and Morzenti, Angelo},
  year={2013},
  publisher={Springer}
}

@inproceedings{hoschele2016mining,
  author       = {Matthias H{\"{o}}schele and
                  Andreas Zeller},
  title        = {Mining input grammars from dynamic taints},
  booktitle    = {Proc. 31st {IEEE/ACM} International Conference on Automated
                  Software Engineering, {(ASE)} 2016},
  pages        = {720--725},
  publisher    = {{ACM}},
  year         = {2016},
  month = sep,
  url          = {https://doi.org/10.1145/2970276.2970321},
  doi          = {10.1145/2970276.2970321},
}

@article{angluin1988queries,
  title={Queries and Concept Learning},
  author={Angluin, Dana},
  journal={Machine Learning},
  volume={2},
  number={4},
  pages={319--342},
  year={1988},
  publisher={Kluwer Academic Publishers Norwell, MA, USA}
}

@article{angluin1987learning,
  title={Learning regular sets from queries and counterexamples},
  author={Angluin, Dana},
  journal={Information and computation},
  volume={75},
  pages={87--106},
  year={1987},
}

@article{angluin1981note,
  title={A note on the number of queries needed to identify regular languages},
  author={Angluin, Dana},
  journal={Information and control},
  volume={51},
  number={1},
  pages={76--87},
  year={1981},
  publisher={Elsevier}
}

@inproceedings{caballero2007polyglot,
  title={Polyglot: Automatic extraction of protocol message format using dynamic binary analysis},
  author={Caballero, Juan and Yin, Heng and Liang, Zhenkai and Song, Dawn},
  booktitle={Proc. 14th ACM conference on Computer and communications security},
  pages={317--329},
  year={2007}
}

@article{sassaman2013security,
  title={Security applications of formal language theory},
  author={Sassaman, Len and Patterson, Meredith L and Bratus, Sergey and Locasto, Michael E},
  journal={IEEE Systems Journal},
  volume={7},
  number={3},
  pages={489--500},
  year={2013},
  publisher={IEEE}
}

@article{lammel2001semi,
  title={Semi-automatic grammar recovery},
  author={L{\"a}mmel, Ralf and Verhoef, Chris},
  journal={Software: Practice and Experience},
  volume={31},
  number={15},
  pages={1395--1438},
  year={2001},
  publisher={Wiley Online Library}
}

@inproceedings{angluin1991won,
  title={When won't membership queries help?},
  author={Angluin, Dana and Kharitonov, Michael},
  booktitle={Proc. 23rd annual ACM symposium on Theory of computing},
  pages={444--454},
  year={1991}
}

@inproceedings{yang2011finding,
  title={Finding and understanding bugs in C compilers},
  author={Yang, Xuejun and Chen, Yang and Eide, Eric and Regehr, John},
  booktitle={Proc. 32nd ACM SIGPLAN conference on Programming language design and implementation},
  pages={283--294},
  year={2011}
}
\end{document}